\documentclass[prd,aps,oamsmath,amssymb,nofootinbib,preprintnumbers]{revtex4}
\pdfoutput=1
\usepackage{graphicx}
\usepackage{dcolumn}
\usepackage{bm}
\usepackage{amsmath}
\usepackage{amsthm}

\usepackage{amsfonts}
\usepackage{euscript,bbm}
\usepackage{ifthen}
\usepackage{psfrag}
\usepackage{slashed}
\usepackage{hyperref}
\usepackage{textcomp,gensymb}
\usepackage{calc}
\usepackage{color}
\usepackage{wasysym}

\newcommand{\uI}{\ensuremath{{}^{235}{\rm U}}}
\newcommand{\uII}{\ensuremath{{}^{238}{\rm U}}}
\newcommand{\uIII}{\ensuremath{{}^{239}{\rm U}}}
\newcommand{\puI}{\ensuremath{{}^{239}{\rm Pu}}}

\newcommand{\np}{\ensuremath{{}^{239}{\rm Np}}}
\newcommand{\ger}{\ensuremath{{\rm Ge}}}
\newcommand{\gern}[1]{\ensuremath{{}^{#1}{\rm Ge}}}
\newcommand{\si}{\ensuremath{{\rm Si}}}
\newcommand{\cns}{CE\ensuremath{\nu}NS}

\newcommand{\be}{\begin{equation}}
\newcommand{\ee}{\end{equation}}
\newcommand{\bea}{\begin{eqnarray}}
\newcommand{\eea}{\end{eqnarray}}

\begin{document}

\begin{flushright}
MI-TH-1540
\end{flushright}

\title{Sensitivity to oscillation with a sterile fourth generation neutrino from\\ultra-low threshold neutrino-nucleus coherent scattering}

\author{Bhaskar Dutta$^{1}$}
\author{Yu Gao$^{1}$}
\author{Andrew Kubik$^{1}$}
\author{Rupak Mahapatra$^{1}$}
\author{Nader Mirabolfathi$^{1}$}
\author{Louis E. Strigari$^{1}$}
\author{Joel W. Walker$^{2}$}
\affiliation{$^{1}$~Mitchell Institute for Fundamental Physics and Astronomy, \\
Department of Physics and Astronomy, Texas A\&M University, College Station, TX 77843-4242, USA}
\affiliation{$^2$~Department of Physics, Sam Houston State University, Huntsville, TX 77341, USA}

\begin{abstract}
We discuss prospects for probing short-range sterile neutrino oscillation using
neutrino-nucleus coherent scattering with ultra-low energy ($\sim 10$~eV - 100~eV)
recoil threshold cryogenic Ge detectors.
The analysis is performed in the context of a specific and contemporary reactor-based experimental
proposal, developed in cooperation with the Nuclear Science Center at Texas A\&M University,
and references developing technology based upon economical and scalable detector arrays.
The baseline of the experiment is substantially shorter than existing measurements,
as near as about 2~meters from the reactor core, and is moreover variable, extending continuously
up to a range of about 10~meters. This proximity and variety combine to provide extraordinary
sensitivity to a wide spectrum of oscillation scales, while facilitating the tidy cancellation
of leading systematic uncertainties in the reactor source and environment.
With 100~eV sensitivity, for exposures on the order of 200~kg$\cdot$y, we project an estimated
sensitivity to first/fourth neutrino oscillation with a mass gap $\Delta m^2 \sim 1 \, {\rm eV}^2$ at an
amplitude $\sin^2 2\theta \sim 10^{-1}$, or $\Delta m^2 \sim 0.2 \, {\rm eV}^2$ at unit amplitude.
Larger exposures, around 5,000~kg$\cdot$y, together with 10~eV sensitivity are capable 
of probing more than an additional order of magnitude in amplitude. 
\end{abstract}

\maketitle

\section{Introduction} 
\label{sec:intro}
Several recent short baseline neutrino experiments hint at the presence of additional neutrinos beyond
the three active components in the Standard Model (SM). The radioactive source experiments of the
GALLEX~\cite{Kaether:2010ag} and SAGE~\cite{Abdurashitov:2009tn} Solar neutrino detectors have
found indications of a deficit of electron neutrinos~\cite{Giunti:2006bj,Giunti:2010zu}.
Independently, very short baseline neutrino experiments with distances of $< 100$ m find evidence for a
deficit of electron anti-neutrinos~\cite{Mention:2011rk}. At least one additional sterile neutrino
with a mass splitting of $\Delta m^2_{14} \sim 1$ eV$^2$ and an approximate 10\% admixture between the
electron flavor neutrino and the new sterile neutrino can potentially accommodate both of these results.
More sophisticated proposals include the introduction of a hidden twin three-generation neutrino sector~\cite{Bai:2015ztj},
which would hypothetically feature the testable short-baseline appearance of SM neutrinos from decays in flight.
In addition, anomalies in data collected by the LSND and MiniBooNE~\cite{Aguilar:2001ty,AguilarArevalo:2010wv,Aguilar-Arevalo:2013pmq}
collaborations may likewise be explained by the invocation of sterile neutrinos.

Theoretical models have been developed to incorporate neutrinos at the $\sim 1$ eV mass scale. The
strongest constraints on these models come from cosmology, in particular the constraints on the number of
additional light degrees of freedom from Big Bang Nucleosynthesis (BBN)~\cite{Steigman:2012ve} and
Planck measurements~\cite{Ade:2013zuv}. Though these constraints severely limit the theoretical
models that have been considered, they can be evaded by introducing a coupling between a vector boson with
$\sim$ MeV mass scale to the sterile
neutrinos~\cite{Hannestad:2013ana,Dasgupta:2013zpn,Ng:2014pca}. In addition to cosmology, Solar
neutrinos can test for the presence of a 4th generation sterile neutrino~\cite{Palazzo:2011rj}.

Dedicated experiments are under development to confirm or rule out the existence of an additional sterile
neutrino. These experiments are designed to detect electron anti-neutrinos on a baseline $\sim 1-20$~m,
which is characteristic of the hypothetical sterile neutrino oscillation length.
Ref.~\cite{Ashenfelter:2013oaa} discusses the prospects for using scintillators to detect electron
anti-neutrinos via inverse beta decay in close proximity to a reactor core, while
Refs.~\cite{Dwyer:2011xs, Gao:2013tha, Borexino:2013xxa} discuss the prospects for using a
radioactive source in close proximity to a scintillator. Ref.~\cite{Drukier:1983gj} first
suggested the possibility of probing sterile neutrino oscillations through coherent elastic
neutrino-nucleus scattering (\cns{}). Ref.~\cite{Anderson:2012pn} has recently updated this
notion, in the specific context of low-energy pion and muon decay-at-rest neutrino sources.

In this paper we discuss the prospects for using the~\cns{} channel and
a new experimental design to search for sterile neutrinos.
In spite of the large predicted cross section, \cns{} has yet to be detected,
primarily because the state of the art of detector technology to this date has
been unable to deliver sufficiently low threshold sensitivity to register deposition
of the kinetic energy of the heavy recoiling nucleus.
In particular, experimental programs that have previously discussed the prospects
for detecting \cns{} from nuclear reactors~\cite{Wong:2003ht}
have typically referenced nuclear recoil thresholds at the keV scale or greater.

We project experimental sensitivities referencing the application of ultra-low threshold (as low as
10~eV) \ger{} and/or~\si{} detectors~\cite{Mirabolfathi:2015pha} to the measurement of~\cns{}.
We consider first the more conservative recoil energy threshold of 100~eV for near-term phase-1
experiments. There are many detector technologies that are rapidly approaching the 100~eV energy threshold in the
current generation of dark matter search experiments~\cite{Agnese:2015nto, deMelloNeto:2015mca, Aalseth:2012if},
and we project a target of this order to be feasible for first generation data taking, as further
justified in Section~\ref{sec:errors}.
For high-mass long-exposure runs, such as a ton-scale 5-year follow up experiment, we
restrict analysis to the more speculative 10~eV energy threshold scenario, which will require
substantial further research and development toward the goal of single electron resolution
phonon-mediated ionization detectors~\cite{Mirabolfathi:2015pha}. 
Our first-generation experimental motivation is, however, direct and imminent, referencing the incremental
evolution of in-hand technology based upon economical and modularly scalable detector arrays
in conjunction with a research reactor site capable of providing large flux rates and enabling
distances-from-core as near as one meter~\footnote{
Very near approach of the mobile reactor core (the active region of which is an approximately $0.35$~m
cube with several additional centimeters of housing material at each side boundary) to the
thermal column identified for housing the planned experiment is presently obstructed by
a $\sim 0.7$~m graphite block, the shielding benefits of which may be sub-optimal to the current application.
Pending detachment of this block, it would become physically possible to locate the center of the reactor
(closer than) one meter from the center of a Germanium or Silicon detector array.}, or between two
and three meters after allowing for the additional depth of shielding materials suggested by detailed simulation.
Specifically, the proposed anti-neutrino source is a
megawatt-class TRIGA-type pool reactor stocked with low-enriched ($\sim 20\%$) \uI{}, which is
administrated by the Nuclear Science Center at Texas A\&M University (TAMU). This adjacency
geometrically enhances the neutrino flux to within about an order of that typifying experiments at
a 30~m baseline from a gigawatt-class power reactor source.  Additionally, it provides an experimental
length scale and median energy that are naturally calibrated for probing some of the most interesting
regions of the $\Delta m^2_{14}$ parameter space. Moreover, the core of this reactor design is mobile,
readily facilitating the collection of data at several propagation lengths, which improves sensitivity
to the oscillatory spatial character of the signal and induces a cancellation of leading systematic error
components. Additional background on the TAMU research nuclear reactor facility (including isotopic
fuel ratios and the computation of the expected anti-neutrino flux), the fabrication and performance of
ultra-low threshold \ger{} and \si{} nuclear recoil detectors, and the physics applications of
neutrino-atom scattering are provided in Refs.~\cite{Dutta:2015vwa,Mirabolfathi:2015pha}, as well
as the references therein.  Note, in particular, that most research reactors are not continuously operated,
although there is no technical reason which prevents this, resources permitting; when referencing time,
we refer to reactor live-time.

This paper is organized as follows.
In Section~\ref{sec:cns}, we predict the~\cns{} rate for our planned
detector setup, including the possible presence of an additional sterile neutrino.
In Section~\ref{sec:errors}, we characterize detector technology, experimental backgrounds,
shielding requirements, and sources of systematic error.
In Section~\ref{sec:schedule}, we establish a logarithmic observation baseline schedule with
exposure compensation for the geometric flux dilution with distance.
In Section~\ref{sec:sensitivity}, we estimate statistical sensitivity of the proposed experiment to
$\Delta m^2_{14}$ and $\sin^2 2\theta_{14}$ for various experimental baselines $L$. 
In Section~\ref{sec:statistics}, we establish a procedure in semi-closed form
for quantifying the statistical preference of an oscillation hypothesis in the presence of data,
and demonstrate the cancellation of leading systematic uncertainties.
In Section~\ref{sec:statistics2}, we describe a higher-order statistical method, which goes beyond 
certain approximations and simplifications made in the prior description.
In Section~\ref{sec:conclusions}, we summarize and present our conclusions.

\section{Scattering rate predictions including sterile neutrinos} 
\label{sec:cns}
In this section we theoretically establish the~\cns{} rate at the envisioned detector setup.
We predict both the SM rate and the rate in the presence of a hypothetical additional sterile neutrino.

The coherent elastic nuclear scattering of neutrinos (with sufficiently low energy, typically $\sim$ MeV)
is a long-standing prediction of the Standard Model that has been theoretically
well-studied~\cite{Freedman:1973yd}.
The differential cross-section for SM scattering of a neutrino with energy $E_\nu$ from a target
particle of mass $M$ and kinetic recoil $E_{\rm R}$ is, in terms of the applicable vector $q_V \equiv q_L+q_R$
and axial $q_A \equiv q_L-q_R$ charges,
\be
\frac{d \sigma}{ d E_{\rm R}} = \frac{G_F^2 M}{2\pi}\bigg[\, (q_V+q_A)^2
+ (q_V-q_A)^2\left( 1-\frac{E_{\rm R}}{E_\nu} \right)^2 - (q_V^2-q_A^2)\frac{M E_{\rm R}}{E_\nu^2} \bigg].
\label{eq:dsdtde}
\ee
For anti-neutrino scattering there is a relative negative phase between $(q_V,q_A)$ associated with the parity-flip,
which is accommodated by the prescription $(q_V,q_A) \equiv (T_3-2Q\sin^2\theta_W,-T_3)$.
For coherent nuclear scattering, these terms should be summed over the quark content of protons and neutrons, and either
multiplied by the respective counts $(Z,N)$ of each (in the vector case) or multiplied by the respective
differential counts $(Z^+-Z^-,N^+-N^-)$ of up and down spins (in the axial case)~\cite{Barranco:2005yy}.
This sum over nuclear constituents at the coupling level, prior to squaring in the amplitude, is the
essence of the nuclear coherency boost.

In order to compute the cumulative expected Standard Model anti-neutrino scattering rate (cf. Ref.~\cite{Barranco:2005yy}),
it is necessary to integrate in the region of the $E_\nu$ vs. $E_{\rm R}$ plane that is
above $E_\nu > E_\nu^{\rm min} \equiv (E_{\rm R}+\sqrt{2 M E_{\rm R} + E_{\rm R}^2})/2$, which is the
minimal neutrino energy (i.e. the inversion of the expression for
the maximum recoil $E_{\rm R}^{\rm max} \equiv 2E_\nu^2/(M+2E_\nu)$ achievable
in a collision with no glancing component) required to trigger a given recoil,
and between the $i^{\rm th}$ binned boundary pair $E_{\rm R}^{i\downarrow} < E_{\rm R} < E_{\rm R}^{i\uparrow}$
of the detector recoil.  The integrand is a product of the previously described differential cross section and the
normalized anti-neutrino energy spectral distribution $\phi(E_\nu) \equiv dN_\nu/dE_\nu \div N_\nu$, as well as
the incident anti-neutrino flux $\Phi$, the detector mass $M$, and the exposure time $T$.
The expected number $N_{\rm Exp}^{i,n}$ of \cns{} scattering events in a detector of composite mass
$M_{\rm Det}$ from nuclei with mass $M$ in a recoil energy bin $E_{\rm R}^i$ for an exposure $T_n$ at a
distance $L_n$ is then
\be
N_{\rm Exp}^{i,n} \,=\,
\Phi_0 \times
T_n \times
\frac{L_0^2}{L_n^2} \times
\frac{M_{\rm Det}}{M} \times
\int_{E_\nu^{\rm min}(E_{\rm R}^{i\downarrow})}^{\infty} \hspace{-28pt} dE_\nu \,\, \phi(E_\nu) \,
\int_{E_{\rm R}^{i\downarrow}}^{\min\{ E_{\rm R}^{i\uparrow} , E_{\rm R}^{\rm max}(E_\nu) \}}
\hspace{-24pt}
dE_{\rm R} \,\, \frac{d\sigma}{dE_{\rm R}} (E_\nu,E_{\rm R}) \, .
\label{eq:nexp}
\ee
Nuclear scattering will generally be dominated by the vector charge, and in the limit of
vanishing axial charge the residual functional dependence $1-M E_{\rm R}/2E_\nu^2$ interpolates between
a large cross-section at zero recoil and a vanishing cross-section at kinematic cut-off.
The large mass-denominator in $E_{\rm R}^{\rm max}$ highlights the
necessity of ultra-low threshold detectors for observation of the heavily boosted \cns{} feature.
Specifically, a typical MeV-scale incident neutrino will be downgraded by almost 5 orders of magnitude
in the recoil (in the most kinematically favorable case) with a 100~GeV order nucleus, to around 20~eV.
In more detail, we calculate that, in order to capture about half of the scattering from fission neutrinos
with a mean energy of 1.5~MeV, a detector threshold around 50, 20~eV is required in \si, \ger.
In this energy regime \cns{} scattering dominates over electron cloud scattering, by a factor
of about 500 at 10~eV, and about 30 at 500~eV; the expected scattering rates become comparable
above about 1~keV, where the available kinematic recoil of a
nuclear mass from MeV neutrino scattering approaches kinematic freeze-out.
For recoil thresholds of 10 or 100~eV, the minimal neutrino energy required to trigger a detection
in \ger{} is around 0.6 or 1.9~MeV.  This is potentially of note when considering the impact of
additional low-energy neutrinos from secondary $\beta$-processes induced by reactor neutrons and gammas.

For concreteness, we have adopted the anti-neutrino energy spectral distribution $\phi(E_\nu)$ of Ref.~\cite{Schreckenbach:1985ep}
to model reactor neutrinos with energies $E_\nu$ above 2~MeV. This spectrum has been extrapolated from the correlated
observation of electrons emitted by the fission products of \uI{} irradiated with thermal neutrons.
The flux is down by about a factor of 20 at 5~MeV, and around six magnitude orders at 10~MeV.
Below 2~MeV (the threshold for inverse $\beta$-decay $\overline{\nu}_e+p \to e^+ + n$ is $E_\nu > 1.8$~MeV)
there is no experimental data, and we employ the theoretically established curve of Ref.~\cite{Kopeikin:2012zz}.
For reference, the intrinsic anti-neutrino production rate of the TAMU research reactor is
approximately $1.9\times10^{17}$ per second, yielding a flux at a mean distance-from-core $L_0$ corresponding to 1~m
of $\Phi_0 \simeq 1.5 \times 10^{12}~[{\rm s}\cdot{\rm cm}^{2}]^{-1}$ (dividing by $4\pi \times 100^2 \sim 10^5$).

All plotted and tabulated numerical calculations in this document are performed in the context of a Germanium target,
averaging over natural isotopic abundances, where the leading contribution (at just over one third) is from
\gern{74}, followed by \gern{72} and \gern{70} (at around one quarter and one fifth respectively),
as well as \gern{73} and \gern{76} (at less than one tenth each).
Typical values of $N_{\rm Exp}^{i,n}$ for \ger{} in the proposed experimental configuration are provided
in TABLE~\ref{tab:expected}, normalized to an exposure of one year with a one kilogram detector at one meter from core.
Keeping in mind that the practical separation of detector and reactor centers may typically be somewhat larger,
it is convenient here to provide baseline signal rates normalized to one meter for numerical simplicity
of the rescaling law from unit length to other experimental baselines.
This is of particular relevance for an oscillation experiment,
wherein it is advantageous to probe several different length scales.
The energy binning adopted is tuned here to achieve
an approximately equitable distribution of scattering events. It is
consistent with the expected detector recoil energy resolution of approximately 10\%,
down to a demonstrated (projected) absolute RMS resolution of around 7 (2)~${\rm eV}_{ee}$~\cite{Mirabolfathi:2015pha}.
Moreover, it produces a manageable collection of curves that are adequately spaced so as to be
structurally distinguishable and statistically well-populated,
yet sufficiently finely grained so as to preserve features of the underlying continuum.
Note, however, that the specific bin boundaries indicated in TABLE~\ref{tab:expected} are purely illustrative,
insomuch as any practical binning scheme will ultimately be very sensitive to the future
precision calibration of low-energy nuclear responses, as described further in Section~\ref{sec:errors}.
In the case of a recoil sensitivity threshold substantially above 10~eV, the inapplicable bins are discarded.

\bgroup
\def\arraystretch{1.5}
\setlength\tabcolsep{8pt}
\begin{table}[h!]
   \begin{center}
     \caption{The SM expected count of \cns{} scattering events $N^i_{\rm Exp}$ per nuclear kinetic recoil
	energy $E^i_{\rm R}$ bin $i$
 	in 1~kg of \ger{} at a mean distance from core of 1~m for an integrated exposure of 1~[y].
	The event rate drops precipitously above 1~keV. The presented binning is merely suggestive,
and will ultimately be strongly dependent upon precise calibration of the nuclear response function
at low energies.} \label{tab:expected}
   \vspace{8pt}
     \begin{tabular}{ c | c c c c c c c c c }
       \hline
	$E^i_{\rm R}~{\rm eV}$ & 10-20 & 20-35 & 35-55 & 55-80 & 80-110 & 110-155 & 155-220 & 220-350 & 350-1000 \\ 
       \hline
	$N^i_{\rm Exp}$ & 853 & 961 & 962 & 911 & 818 & 874 & 820 & 862 & 782 \\
       \hline
    \end{tabular}
  \end{center}
\end{table}
\egroup

We now consider the modification of these rates in the presence of a fourth sterile neutrino flavor, with a
mass gap from the SM triplet on the order of $\Delta m^2_{14} \simeq 1~{\rm eV}^2$. Our targeted
experimental baseline is on the order of a few to several meters (starting from the minimal length
consistent with shielding requirements, e.g. 2-3~m, and extending to the vicinity of 10~m),
such that SM oscillation to mu and/or tau flavors (which occur over much longer distances) is entirely decoupled.
In this case, the probability $P_{(\alpha\to\beta)}$ for oscillation between the two decoupled
neutrino flavors $(\alpha,\beta)$ is
\begin{equation}
P_{(\alpha\to\beta)} = \sin^2 \big[ 2 \theta \big] \times \sin^2 \big[ \frac{ \Delta m^2 L }{4 E_\nu }
\big] \,.
\label{eq:oscprob}
\end{equation}
The wavelength associated with this propagation may be expressed as
\begin{equation}
\lambda = 4.97~[{\rm m}] \times
\left\{\frac{E_\nu}{1~[{\rm MeV}]}\right\} \times
\left\{\frac{1~{\rm eV}^2}{\Delta m^2}\right\} \,. 
\label{eq:wavelength}
\end{equation}

\begin{figure}[ht]
\centering
\hspace{-5pt}
\includegraphics[width=5.0in]{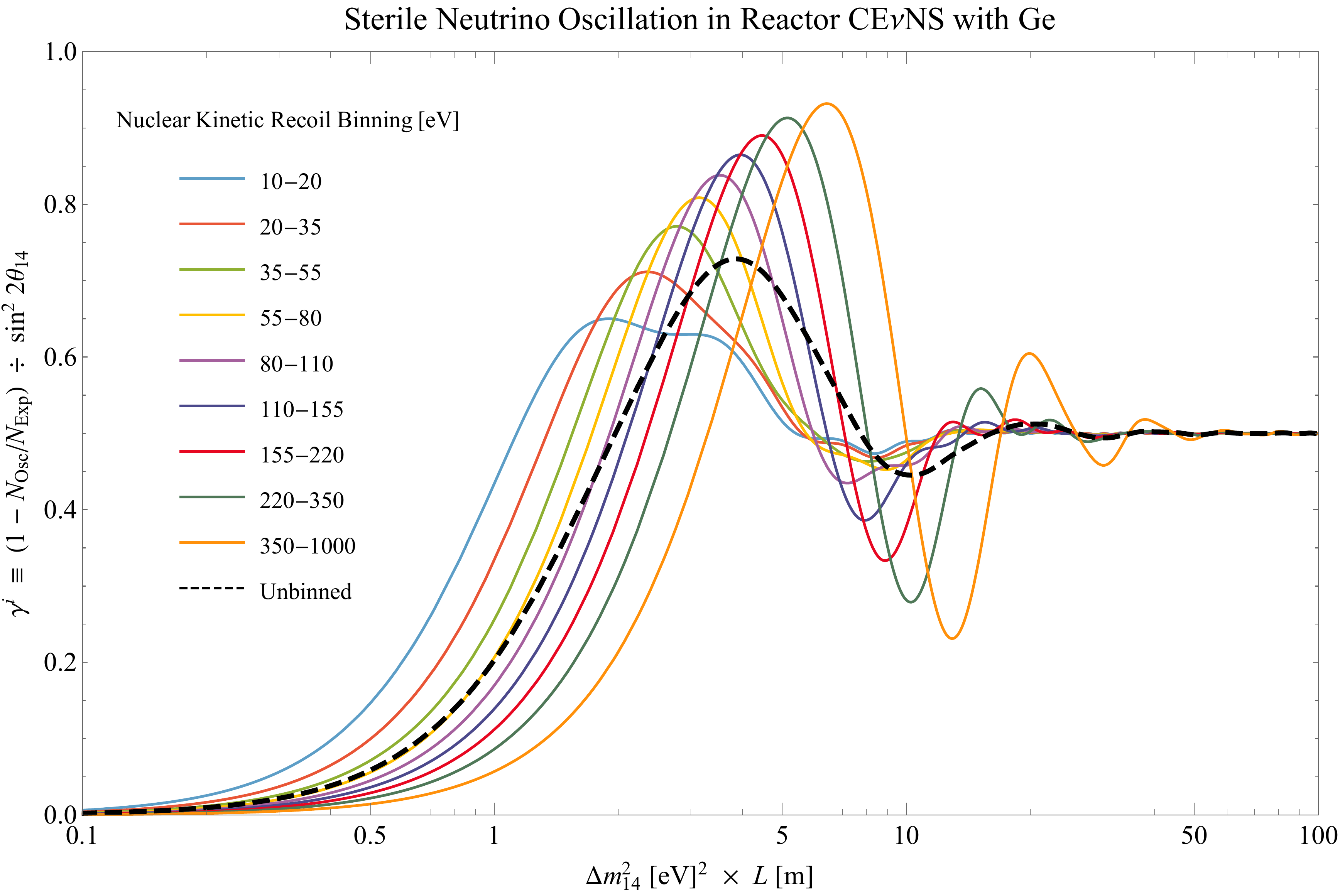}
\caption{
\footnotesize Fractional deviation from expected SM \cns{} event rates in \ger{}
due to electron anti-neutrino oscillation with a sterile fourth flavor.
Separate basis curves $\gamma^i$ are displayed for nine binned windows in the nuclear
kinetic recoil energy deposition. A tenth (bold, dashed) curve
demonstrates the cumulative unbinned event deviation over all recoils above 10~eV.
For a given energy band, values approaching 1 indicate more full depletion relative to the SM,
whereas values nearer to zero indicate more marginal depletion.
The vertical axis is inversely rescaled by the amplitude factor $\sin^2 2 \theta_{14}$,
and predictions for any targeted amplitude may be immediately read off by
replacing the upper limit (1.0) with the applicable value. The horizontal
axis indicates the product of the distance from core in meter units
and the mass-square difference $\Delta m_{14}^2$ in units of ${\rm eV}^2$.
For example, at a gap of $1~{\rm eV}^2$, this axis is read
literally in meters, whereas the outer scale bound at 100 would correspond instead
to 10~m for $\Delta m_{14}^2 = 10~{\rm eV}^2$.  The presented binning is merely suggestive,
and will ultimately be strongly dependent upon precise calibration of the nuclear response function
at low energies.}
\label{fig:oscillation}
\end{figure}

It is useful to define a dimensionless quantity $\gamma_i$ that represents depletion
in the observed \cns{} scattering rate $N^i_{\rm Osc}$
relative to the SM expectation $N^i_{\rm Exp}$
due to oscillation with a sterile fourth generation neutrino
(for some binning index $i$ in the energy width and/or detector location).
\begin{equation}
\gamma_i(\Delta m^2_{14} L) \,\equiv\, \frac{1\,-\,(N^i_{\rm Osc}/N^i_{\rm Exp})}{\sin^2 2\theta_{14}}. 
\label{eq:gamma}
\end{equation}
We may then explicitly establish the convolved oscillation shape
functionals $\gamma_i(\Delta m^2_{14} L)$ defined in Eq.~(\ref{eq:gamma}).
\be
\gamma_i(\Delta m^2_{14} L) \,=\, \left\langle\sin^2\big[\frac{\Delta m^2_{14} L}{4 E_\nu}\big]\right\rangle_{\hspace{-3pt}E_\nu} \equiv\,
\iint dE_\nu \,d\sigma \,\phi \times \sin^2\big[\frac{\Delta m^2_{14} L}{4 E_\nu}\big] \,\div\, 
\iint dE_\nu \,d\sigma \,\phi
\label{eq:gammafull}
\ee
Integration bounds and function dependencies have been suppressed in Eq.~(\ref{eq:gammafull}), but are identical
to those expressed explicitly in Eq.~(\ref{eq:nexp}).
In particular, $d\sigma \equiv dE_{\rm R} \,\, \frac{d\sigma}{dE_{\rm R}} (E_\nu,E_{\rm R})$ represents
a differential in the cross-section to be integrated over recoil energy, and $\phi \equiv \phi(E_\nu)$
represents the reactor anti-neutrino source spectrum.
There is no residual dependence in the $\gamma_i$
upon the exposure time, source flux normalization, or detector mass.
Once computed (for example, by sampling at fixed increments in the product $\Delta m^2 L$,
and interpolating to a continuous numerical function), 
The dimensionless $\gamma_i(\Delta m^2_{14} L)$ wholly encapsulate all
theoretical aspects of a putative oscillation signal's length and mass scale dependence, including
implicit dependence upon the anti-neutrino source, and the physics of the recoil detection mechanism.

FIG.~(\ref{fig:oscillation}) plots $\gamma_i$, employing the TABLE~\ref{tab:expected} binning,
after convolution of a suitable reactor source spectrum
with the \cns{} detector response in \ger{}, as a continuous function of the
product of the mass gap $\Delta m_{14}^2$ and the distance from core $L$.
The coefficients $L$ (horizontal axis) and $1/\sin^2 2\theta_{14}$ (vertical axis)
are adopted in order to make the intrinsic scale invariance of the elementary event profiles manifest.
As such, the depicted curves are not associated with any particular benchmark point in the model space,
but are instead directly applicable to the full range of targeted parameters by simple
renormalization of the axes with respect to the amplitude and phase of Eq.~(\ref{eq:oscprob}).

At very small values of $\Delta m^2_{14} L$, i.e. at lengths $L$
much smaller than the relevant Eq.~(\ref{eq:wavelength})
wavelength $\lambda$, the oscillation fraction will be inadequate for observation. Conversely, at excessively
large values of $\Delta m^2_{14} L$, where the propagation length represents several half-wavelengths,
dispersion will wash out all discernible features, $(\gamma_i \Rightarrow 1/2)$.
The recoil energy binning is observed to partially mitigate
the expected wavelength dispersion at both long and short distance scales,
as is evident in comparison with the unbinned integration of all recoils.
By resolving individual binned response curves, the window of sensitivity to oscillation in the
mass gap $m_{14}^2$ parameter is widened by almost an order of magnitude at fixed $L$. By physically
modulating the experimental baseline $L$ (as facilitated by ready mobility of the reactor core at the
proposed experimental site), this window may be extended even further.

\section{Characterization of Detectors, Backgrounds and Experimental Uncertainties}
\label{sec:errors}

The march toward detector technology sensitive to nuclear recoil as soft as 100~eV is proceeding
rapidly~\cite{Agnese:2015nto, deMelloNeto:2015mca, Aalseth:2012if}.
In particular, electron recoil thresholds of 56~${\rm eV}_{ee}$~\cite{Mirabolfathi:2015pha}, corresponding under the
assumption of Lindhard scaling~\cite{Lindhard} to a nuclear recoil threshold as low as 250~eV,
have already been actively demonstrated by CDMSlite at Soudan~\cite{Agnese:2015nto}.
Continuous research and development effort is being exerted toward advance on this front,
with fabrication and testing of improved design prototypes currently at various stages of progress.
TABLE~\ref{tab:detectors} summarizes vital characteristics of the first-generation detectors envisioned
for deployment in the described experiment, including recoil threshold, net mass, and energy resolution.
Specifically, an RMS resolution of 7~${\rm eV}_{ee}$ has now been locally demonstrated in a wafer of
1~cm thickness~\cite{Mirabolfathi:2015pha}), and scaling to 3.3~cm detectors with the same demonstrated bias voltage
will ostensibly provide for a resolution of $7\div3.3 \sim 2~{\rm eV}_{ee}$.
This corresponds to a projected electron-equivalent recoil threshold of approximately
10~${\rm eV}_{\rm ee}$, which scales under various extrapolations of the Lindhard factor to
a threshold for nuclear recoils that is between 50 and 100~eV, allowing for a wide uncertainty
pending the direct calibration of this low-energy regime.

\bgroup
\def\arraystretch{1.5}
\setlength\tabcolsep{8pt}
\begin{table}[h!]
   \begin{center}
     \caption{We summarize expected characteristics of the first generation germanium detectors to be
	available for application to the described reactor sterile neutrino oscillation experiment.
	Demonstrated electron recoil resolution~\cite{Mirabolfathi:2015pha} is projected onto
	the target wafer thickness, and extrapolated under the assumption of Lindhard~\cite{Lindhard}
	scaling to a likely range of energy thresholds for the sensitivity to nuclear recoils.
	We allow for a conservative error factor of 2~in this projection, since the direct calibration of nuclear
	recoils in this energy regime remains to be established.}
	\label{tab:detectors}
   \vspace{8pt}
     \begin{tabular}{ | c | c | }
       \hline
	Dimensions ($\diameter \times {\rm h}$) & $100~{\rm mm} \times 33~{\rm mm}$ \\
       \hline
	Mass & 1.4~kg \\
       \hline
	Resolution $\sigma~(E=0)$ & 2~${\rm eV}_{\rm ee}$ \\
       \hline
	Electron $E_R^{\rm min}$ &  10~${\rm eV}_{\rm ee}$ \\
       \hline
	Lindhard Factor & 0.1~\textendash~0.2 \\
       \hline
	Projected Nuclear $E_R^{\rm min}$ & 50~\textendash~100~eV \\ 
       \hline
    \end{tabular}
  \end{center}
\end{table}
\egroup

Reaching the lowest 10~eV recoil sensitivity included in the present theoretical treatment
will require further technological advances, although this goal would seem to be a plausible
one with regards to basic underlying physics.  Specifically, it will require the eventual
realization of high-bias phonon-mediated ionization detectors with single-electron sensitivity~\cite{Mirabolfathi:2015pha}.
The Lindhard factor at such low energies, or even at 100~eV, is yet to be established through direct
nuclear recoil energy calibration.  Current generation scattering experiments with a neutron beam
targeted on commercial low-energy Ge detectors are capable of reaching 0.5~keV nuclear recoil thresholds,
extending the work of Ref.~\cite{Jastram}. 
Nuclear recoil calibration at very low energies, down to around 100~eV, is expected to be
performed as part of the SuperCDMS SNOLAB program via the exposure of high-voltage detectors
to either a neutron beam or activated source in an underground test facility.
The ultimate goal is calibration down to the order of 10~eV nuclear recoils.
In this regime, at the edge of single electron production, the practically achievable threshold
will be sensitive to composition of the target material; for example, whereas a 10~eV threshold
might be realistic in \ger{}, a number closer to 20~eV might be more appropriate to a material such as \si{}.
At higher energies, on the order of 100~eV, the technologically limiting factors are less sensitive to such distinctions.

The placement of sensitive experimental apparatus in the near-field of a working nuclear reactor
presents unique challenges associated with characterizing and modulating this radiologically intense setting.
Primary backgrounds inextricably linked with the targeted anti-neutrino flux include prompt gammas and neutrons sourced from the core.
The energy spectrum and production rate of these backgrounds are predicted using a
MCNP~\cite{MCNPX} core model developed at the Nuclear Science Center at Texas A{\&}M University.
Additionally, there are secondary background products derived from interactions with various casing and shielding materials.
Sub-leading radiation levels from the decay of quasi-stable daughter products within the reactor, and from activated structural
materials surrounding experimental cavity, persist even after the primary fission reaction has been powered down.
Further, any surface-level experiment must also contend with muons and muon-induced neutrons from cosmic rays,
as well as ambient gammas.  The rate and spectral shape of detected recoils from such backgrounds must be either measured
directly or estimated using simulation, and a shielding design must be introduced that is capable of reducing these backgrounds
to a ceiling order-commensurate with (or below) that of the expected neutrino recoil rate.

Dual programs have been initiated for simulation in GEANT4~\cite{Agostinelli:2002hh}
of the experimental environment (including prototype shielding) and detector response
to reactor neutron and gamma sources, and for the direct in-situ measurement of the same.
The background shape and rate normalization now compare favorably between these two approaches.
Certain details of the background estimation are provided subsequently in this section.
The iterative process of measurement, simulation, and shielding design continues, and is described
more comprehensively (and also more quantitatively) in a parallel publication with a primarily experimental focus~\cite{Agnolet:2016zir}.
In particular, the interested reader may there review in detail the 
nuclear simulation of neutron and gamma reactor flux rates (FIGS.~4),
measurement of the gamma spectrum at various reactor power settings and core distances (FIGS.~7,8),
comparison of the measured background shape and scaling against environmental simulation (FIGS.~9,10),
and estimates of the shielding-controlled neutron and gamma flux rates in the signal region (FIG.~14).

In support of the present, primarily theoretical, analysis, we will begin from a target specification of the maximal background rate
consistent with sensitivity goals, and then ask whether projected estimates of the corresponding shielding requirements are
compatible with practical design limitations.  Specifically, for observations in the nearest field,
we will stipulate a unified (gamma and neutron) background
rate in the detector of no more than 100~dru (differential rate unit, i.e. events/kg/day/keV)
within the 10 to 1,000~eV region of interest, and will also secondarily consider a reduced background level of 10~dru.
For observations performed at a more extended distance from the reactor, which benefit from additional natural depletion
via reduction of the solid angle and increased line-of-sight integration depth through the pool (and also
possibly a greater depth of applied shielding), but which likewise suffer from geometric reduction in the neutrino
flux, we will target 10 or 1~dru, respectively.
These values are suggested as controllable bounds relative to the presented projection for the \cns{} rate
in \ger{}, which corresponds in turn to about 20~dru at one meter from core, about 5~dru at two meters,
and about 1~dru at five meters.
We have folded the background target into projections for the experimental reach, have projected the amount
of data collection necessary in order to reach a measurement to 1\% statistical resolution of these
backgrounds in the experimental sideband above 1-2~keV, and have clarified the expectations for a shielding
design capable of facilitating this goal.

Prompt gamma and neutron byproducts of fission in the reactor core are radiated with an integrated flux 
comparable to that of the signal neutrinos, i.e. on the order of $10^{12}~[{\rm s}\cdot{\rm cm}^{2}]^{-1}$ at a distance of about one meter.
These backgrounds diminish with distance-from-core (in a shield-dependent manner), but do not participate in the
neutrino oscillation, and thus may potentially inhibit discrimination of the oscillation amplitude $\sin^2 2\theta_{14}$.
Reactor gammas and neutrons are generated with an energy spectrum as given by the MCNP core model,
produced at the face of the reactor core for a given reactor position.  This has allowed for an estimation of
the flux and deposited recoil energy at the detector location, when combined with
a model geometry of the experimental hall (including an accurate description of all materials)
constructed in the GEANT4 framework.
To test the simulation, in-situ measurements of the gamma background were taken with a HPGe (roughly 0.5~kg) detector
and minimal lead shielding, located at the proposed experimental location.  Energy deposition spectra were obtained with the core at
various positions and power outputs, and compared to the prediction of the GEANT4 simulation.  The simulation was found to accurately
predict both the energy spectrum (on the order of 1\% agreement) in the detector, as well as the scaling of the rate with distance from the core to within 10\%.
A preliminary shielding design was then added to the GEANT4 geometry model to asses the background expected in the full experimental
setup.  This shielding included both approximately 1.3~m of borated (5\%) polyethylene as neutron shield and 30~cm of lead as gamma shield.
Results indicate that backgrounds may be controlled to roughly the required order by adopting a shielding design
corresponding to a mean core-to-detector separation of at least about 2~m.
Such shielding was also found to be adequate for reduction of the ambient gamma background, which is much smaller than the reactor counterpart.
A key lesson extracted from the early simulation data is that the interleaving of various shielding layers is essential
to the suppression of backgrounds.  The reasons for this are that each neutron capture will induce emission of a gamma,
and likewise, high energy gamma rays may induce a secondary neutron cascade.  Therefore, the sequential alternation of
shielding materials respectively most effective against neutrons (e.g. poly) or gammas (e.g. lead) may be necessary in order
to prevent the subsequent regeneration of a previously controlled background species. 

Further optimization of the shielding design, including the described interleaving, is now underway.
The described physical baseline of 2~m is apparently compatible with the proposed experimental objectives,
and is presently adopted as a minimal practical separation of the detector and reactor centers.  By increasing this separation, and
back-filling the intervening space with additional shielding materials, it appears possible to further substantially reduce
reactor backgrounds below the specified maximum.  Mobility of the reactor core within its pool environment provides a free mechanism
for variably enhancing early damping of the neutron component without any reconfiguration of the experimental column due
to the considerable efficacy of simple water shielding against this background.
Given that all backgrounds will tend initially to fall much more rapidly with distance in an appropriate shielding material
than the geometric $r^{-2}$ neutrino depletion, and then ultimately level out into a regime of deep suppression,
one generically expects a metric of signal $S$ to background $B$ significance such as $S/\sqrt{S+B}$ to become maximized
for some intermediate distance scale.  As such, it remains possible that the experimental sensitivity to certain phenomena may be elevated
by extending the minimal separation baseline, e.g. to three meters.

The interaction of a final state neutron with the detector will be dominantly in the form of a nuclear recoil,
whereas incident gammas will interact prominently with the atomic electron cloud.  At large recoils, independent
phonon and ionization measurements facilitate an event-by-event discrimination of these two types of interaction.
However, at lower recoil energies, below about 1~keV, convergence of the respective yield curves precludes such an identification.
In this regime, which corresponds to the region of interest for the detection of \cns{}, it may instead be advantageous to employ
a large bias voltage in order to amplify the secondary Luke-Neganov phonon (i.e. phonons shed by drifting charge carriers) signal
far above the energy scale of the original recoil event.  In so doing, the threshold of sensitivity to very soft recoils may
be dramatically extended, with the fundamental energy scale reconstructed from inference of the number of electron-hole pairs
created in the impact, as inferred in turn from the phonon calorimetry, which is proportional to this quanta times the applied voltage.
Incidentally, this also implies that the available energy binning is not strictly arbitrary, but is rather dependent upon the 
stepwise function relating quantized production of electron-hole pairs to recoil energy.  An energy resolution below 10~eV
is possible~\cite{Mirabolfathi:2015pha}, which is consistent with the spirit of the example binning in TABLE~\ref{tab:expected}.
Monitoring of backgrounds will be performed with dedicated detectors, including both scintillation-based detectors placed at strategic
locations within the proposed shielding, and a dedicated CDMS iZip-type detector mounted alongside the neutrino detectors.
These monitoring detectors will provide normalization and discrimination for backgrounds above 1~keV recoil energy,
which will provide a strong constraint on the background in the signal region
when coupled with the simulation described previously.

The signal rate associated with \cns{} in \ger{} is suppressed by a factor of about 200 from 10~eV to
1~keV, and is practically non-existent above 2~keV.  By contrast, the reactor backgrounds extend far above
this scale, and may thus be precisely calibrated by observation of the experimental side-band region where
\cns{} events are unpopulated.  Simulation and preliminary observations suggest that the recoil spectrum
in this regime is essentially flat, as consistent with expectations for low-energy Compton scattering.
It will be necessary to even more carefully model the background shape in the signal region for purposes
of the planned experiment, but it is possible already to make certain important observations.
Statistical calibration of the background normalization to, say, 1\% will require a number
of events on the order of $N \simeq 10^4$ in order to satisfy $\sqrt{N}/N \simeq 0.01$.
Taking, for example, a conservative side-band region extending from 2~keV up to about 10~keV
(in reality, there will be usable normalization data far above this scale)
this resolution would be achievable after approximately just a single running day with a single kilogram
detector at a background rate of 100~dru, and in less than two weeks at 10~dru.  Although this
analysis is quite simplistic, it conveys an essential point: especially in the regime that
backgrounds are large enough to compete with signal, then they are also quite easy to calibrate
for the purpose of generating a data-driven background estimate in the signal region of interest.

In addition to the described reactor backgrounds, there are cosmogenic backgrounds, primarily in the form of
muon-induced neutrons (e.g. via spallation in the concrete reactor pool housing),
which do not decline with distance from core\footnote{
We emphasize here that the experimental plan is to hold the detector arrays and shielding at a fixed location, as the reactor core
is track-mounted and mobile.  This is a significant advantage of the host reactor setting.
Benefits of this architecture include increased mechanical stability of
of the experimental apparatus, and increased geometric stability of each detector's complex relationship with
its surroundings.  Specifically, this carries over to an expectation for immutability of the cosmogenic backgrounds.};
backgrounds of this type are particularly difficult for samples taken with the reactor at the far positions, as their uniform deposition rate at all
detector baselines integrates to a much more substantial fraction of all events as the exposure time $T_n$ is escalated to compensate
for the geometric flux dilution. It is possible, however, to estimate this rate and to make an
appropriate correction to each observation; the correction factor may be modeled computationally, or (better)
extracted from data as a DC signal component that does not conform to a power law or oscillatory profile, or
(best) measured directly during reactor off-cycles (although one must be careful to account for the residual
activity of medium-lived daughter products (e.g. from the neutron capture-induced breeding of \uII{} to \uIII{},
which beta-decays over 2-3 days to \puI{} via \np{}).
In-situ measurements of the muon rate at the experimental site have now been collected,
and show a reduction of roughly a factor of 4 due to the concrete overburden
(about 13 mwe) of the reactor pool wall.  Direct discrimination of the associated (unified) neutron event rate
is furthermore achievable in the same manner previously described for reactor-sourced
neutrons, via implementation of a dedicated detection module with supplementary ionization data capture.
Ultimately, this potentially dangerous background is expected to be quite well controlled relative to an 
intrinsic rate of around a quarter hertz per kilogram.  Beyond the described overburden, 
active veto technology is available with approximately 97\% efficiency,
and (most simply and most importantly) a further thousand-fold reduction is achieved through 
a kinematic cut above the \cns{} threshold $E_{\rm R} \lesssim 1$~keV.

Given the close proximity of source and detector described in this document, it is natural to ask 
when one should expect the point-like approximation of each physically extended object to remain valid. 
In particular, the spatial distribution of active flux sources
(as well as the somewhat more narrowly confined distribution of measurement positions)
will induce additional smearing in the basis-functions presented in FIG.~(\ref{fig:oscillation}).
Such smearing may be expected to manifest
as perturbative corrections in the ratio of the core size, i.e. 0.3~m, over the experimental baseline, which is typically
close to (and may be substantially more than) an order of magnitude larger.  Given the smallness of this parameter, we expect that the
dominant effects along these lines arise instead from the previously modeled
convolution of the oscillation response with the reactor energy spectrum
and the recoil differential cross-section.  Incidentally, such correction terms would break the
scale-invariance manifest in FIG.~(\ref{fig:oscillation}).  Although the baseline projections made in this
document should be insensitive to these details at leading order, it will be important to more completely
simulate such finite-size effects as the described project approaches the data collection phase.

It is worthwhile also to consider several additional sources of systematic experimental uncertainty,
which may impose a ceiling on the expected statistical resolution.  A very important uncertainty exists
with regards to the overall anti-neutrino flux normalization (and to a lesser degree, shape).  Errors are propagated from
imperfect knowledge of the reactor thermal power, and from the extrapolation of this power into the associated anti-neutrino spectrum.
Reactor operators are able to provide precision measurements of
the thermal output power, as well as estimates (based on simulation with
the code {\tt MCNP}~\cite{mcnp}) of the isotopic fuel composition and fission fractions $f_i/F$,
where $f_i$ is the absolute fission rate of species $i$ and $F\equiv\sum f_i$.
Uncertainty estimates on the order of 2-3\% are typical, although it may be possible to reduce
this to around a half of a percent (cf. Ref.~\cite{Cao:2011gb}).  However, it remains to be
clarified precisely how the cited improvements may scale from the context of a commercial power
reactor to that of a TRIGA research reactor.  Calibration of direct recoil observations in the experimental
sideband are also able to independently aid determination of the appropriate normalization factor.

The experimental apparatus is furthermore itself vulnerable at some level to both binary (false or failed
trigger response) and continuous (smearing of the recoil energy resolution) errors, although these
effects are expected to be rather small, with efficiencies rapidly approaching 100\% within the fiducial volume
and just above the recoil threshold.
Millisecond recovery time in the detector modules eliminates substantial danger of event pileup.
Finally, additional (minor) uncertainties may be propagated from errors in input parameters (e.g.
th renormalization group evolved gauge couplings and Weinberg angle or the $Z$-boson mass), limitations in the analysis (e.g.
the omission of electron scattering relative to \cns{} events or imperfect accounting of the dispersive
spatial integration about the mean distance between reactor and detector centers), or even the inadvertent
observation of alternate modes of new physics (e.g. $Z^\prime$ scattering or neutrino magnetic moment scattering).

The leading component of the described systematics is expected to proportionally scale the SM
and oscillated (loss of signal) event rate at all energies and all distances, or to present an additive
background event profile which may be estimated and calibrated in the experimental sideband.
In the search for exotic interaction vertices, it is very useful to combine scattering observations from multiple nuclei,
such as \ger{} and \si{}, in order to cancel systematics and resolve signatures in the differential coupling to protons and neutrons.
In the present case, the unique spatial variation of the signal response provides an intrinsic mechanism for the reduction of
systematics with a single scattering target. In this environment, there is an advantage to employing a dominantly \ger{} based
detector paradigm, as it delivers a threefold advantage in the \cns{} interaction rate per kilogram at low threshold.
To the extent that systematics are primarily attributable to a universal scaling error,
they may be dealt with in a particularly tidy fashion, as will be clarified in Section~\ref{sec:statistics}.
Generalized methods applicable beyond this simplified case are described in Section~\ref{sec:statistics2}.

In summary, the described experimental environment is host to a difficult conglomeration of
particle backgrounds, each presenting their own unique profile and complications. However,
preliminary study indicates that a careful combination of shielding, vetoes, simulation, and calibration
may be sufficient to firstly curtail competing rates within the order of definite projections
for the SM \cns{} rate, and secondly characterize the shape and scale of unified
residual recoil backgrounds and systematics to a precision and accuracy facilitating their subtraction.
It is important here to emphasize that the \cns{} signal is not an inherently rare one;
although it is has thus far eluded experimental detection, technology sufficient to
register the associated soft nuclear recoils, in conjunction with a source presenting
a sufficiently large neutrino flux, will certainly deliver observations of coherent nuclear
scattering with very robust statistics.  Such a setting constitutes an ideal laboratory
for probing deviations from the SM in the neutrino sector.  Specifically, the context
of mixing with a sterile neutrino sector provides a profound handle for resolving new physics embedded within
even an imperfectly characterized background via fitting of an oscillation template
correlated across length scales and across the spectrum of recoil energies, facilitating
the cancellation of leading systematics with a non-oscillatory profile.

\section{Exposure per Baseline Observational Schedule}
\label{sec:schedule}

A rough estimate of the exposure necessary to attain statistical significance may
be constructed as follows. From Eq.~(\ref{eq:oscprob}), the maximal anti-neutrino disappearance
fraction is $\sin^2 2\theta$, assuming extremization of the position-dependent
term (every length $L$ will realize this criterion for certain discretized incident
neutrino energies $E_\nu$). Neglecting systematic effects for the time being,
the statistical significance may be gauged by the ratio of the extremal event deficit
$N_{\rm Exp} \sin^2 2\theta$ over the fluctuation $\sqrt{N_{\rm Exp}}$ in the SM
expected event rate $N_{\rm Exp}$. Integrating across energy bins, the net annual SM \cns{} expectation for
a $M=1$~kg \ger{} detector with 10~eV (100~eV) kinetic recoil sensitivity at a distance
of 1~m from a megawatt nuclear reactor source is approximately $N_{\rm Exp} = 8\times10^3$ ($4\times 10^3$) events.
Setting the significance ratio to unity, and taking into account the dilution of flux
with distance $L^2$ from the core, the minimal exposure time in a given sampling bin is around
\begin{equation}
T_{\rm Min} \simeq 1~[\rm y] \times \frac{1.3\times 10^{-4}}{\sin^4 2\theta} \times
{\left\{\frac{1~[\rm kg]}{M}\right\}} \times {\left\{\frac{L}{1~[\rm m]}\right\}}^2 \,.
\label{eq:mint}
\end{equation} 
For example, adopting the stipulated reactor power with a single kilogram detector,
and taking an oscillation amplitude $\sin^2 2\theta \simeq 0.1$, the minimal integration time
for onset of statistical resolution would be about five days at a distance of $L=1$~m,
or about 120 days at a distance of $L=5$~m. The escalation to a 100~kg detector would
offset a ten-fold reduction in the signal amplitude to $\sin^2 2\theta \simeq 0.01$ with identical
exposure. In case of a 100 eV threshold, the event rate is reduced by about 50\%
and the sensitivity to fast oscillations is reduced by about 30\%.
However, forfeiture of the low-energy spectral data implies a loss of independent
short-distance information, which is potentially more damaging than the simple loss of statistics.

In order to resolve the underlying oscillatory character of the signal, one is clearly motivated to sample multiple points
in the oscillation profile. To some extent, this can be accomplished passively, by sampling multiple neutrino energies $E_\nu$
(or multiple kinetic recoil energies $E^i_{\rm R}$) at a fixed distance.  In this manner, cf. FIG.~(\ref{fig:oscillation}),
regular trends in the event deficit might be resolved within a single vertical constant-$L$ slice of the binned response curves.
However, it is obviously preferable to complement this approach with sample data that is literally extended in space, such
that various candidate signal wavelengths may be probed directly. If this is possible, then independent likelihood optimizations
of the $\sin^2 2\theta_{14}$ and $\Delta m^2_{14}$ parameters within an oscillation template may be extracted from each binned
energy range, and subsequently combined into a unified signal fit and error estimate.

Intuition suggests that the best statistical power in an oscillation template fit will emerge from multiple
samples located commensurately with the signal half-wavelength.
As highlighted in FIG.~(\ref{fig:oscillation}), data samples in the extreme near-field are very poorly suited
for resolving oscillation features, which have not had a sufficient baseline over which to mature.  Conversely,
measurements in the very far-field will have dispersed beyond the level at which isolated features may be resolved.
Equally tempered spacings at $\lambda/4$ or wider experience an elevated danger of coincidentally aligning with the signal troughs.
These observations suggest a logarithmic sample locating schedule, with a local density that separates adjacent measurements
by something like a quarter of their mean distance from core.

The prescription for the location $L_n$ of the $n^{\rm th}$ sample point out of $(N+1)$ is as follows, in
units of the primary $(n=0)$ observation scale $L_0 \equiv 1$, as a function of the scaled
position of the final $(n=N)$ observation at $L_N \equiv L$.
\begin{equation}
L_n \,=\, L^{n/N}
\label{eq:lsamples}
\end{equation}
Near detector samples have a very distinct statistical advantage in terms of the $1/L^2_n$ flux enhancement,
which radically reduces requisite integration times relative to those of far samples.
In order to accrue a matching event count at longer distances, an offsetting quadratic compensation
$T_n \propto L_n^2$ in the exposure time is required.  The constant of proportionality may be established by
constraining the sum over exposures to match a specified net interval $T$,
after which the time $T_n$ per sample location $L_n$ is prescribed as follows.
\begin{equation}
T_n \,=\, T \times L^{2n/N} \times
\left( \frac{L^{2/N} \,-\, 1}{L^{2+2/N} \,-\, 1} \right)
\label{eq:tsamples}
\end{equation}
The increased temporal cost of far detector samples is attenuated by elongation of the sample spacing.
Like the fretting intervals on a guitar, each ``octave'' (doubling of the experimental baseline) will
contain an identical number of sampling locations in this prescription, although it will necessitate
a four-fold increase in the cumulative exposure time to achieve similar statistical significance.
It should be emphasized that the described accommodation provides for the delivery of equal statistical
weight from all sample locations, but it simultaneously escalates the impact of backgrounds,
primarily cosmogenic, that do not abate with distance in the far samples.

\section{Estimation of Expected Sensitivity}
\label{sec:sensitivity}

\begin{figure}[ht]
\centering
\hspace{-5pt}
\includegraphics[width=3.2in]{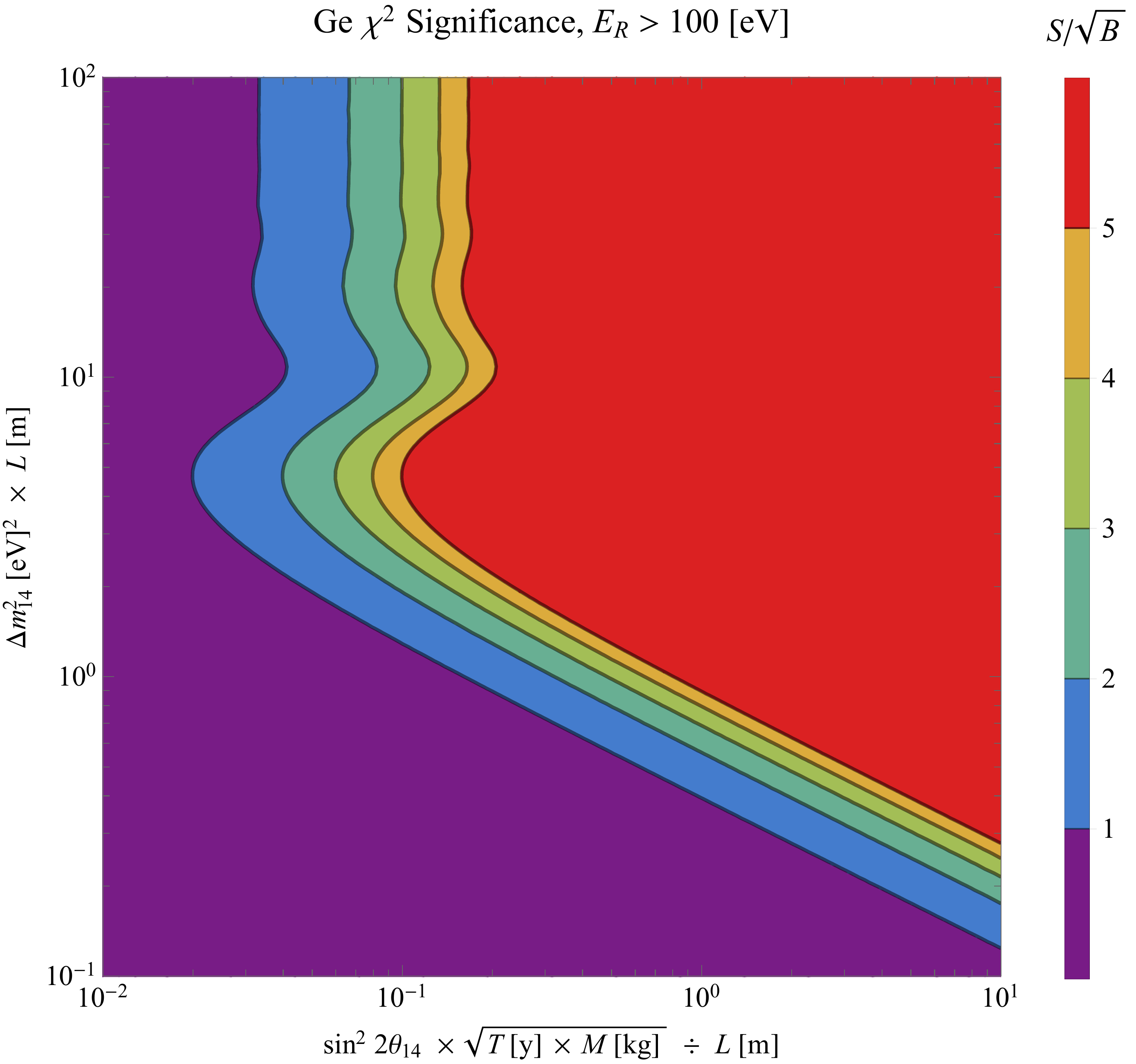}
\includegraphics[width=3.2in]{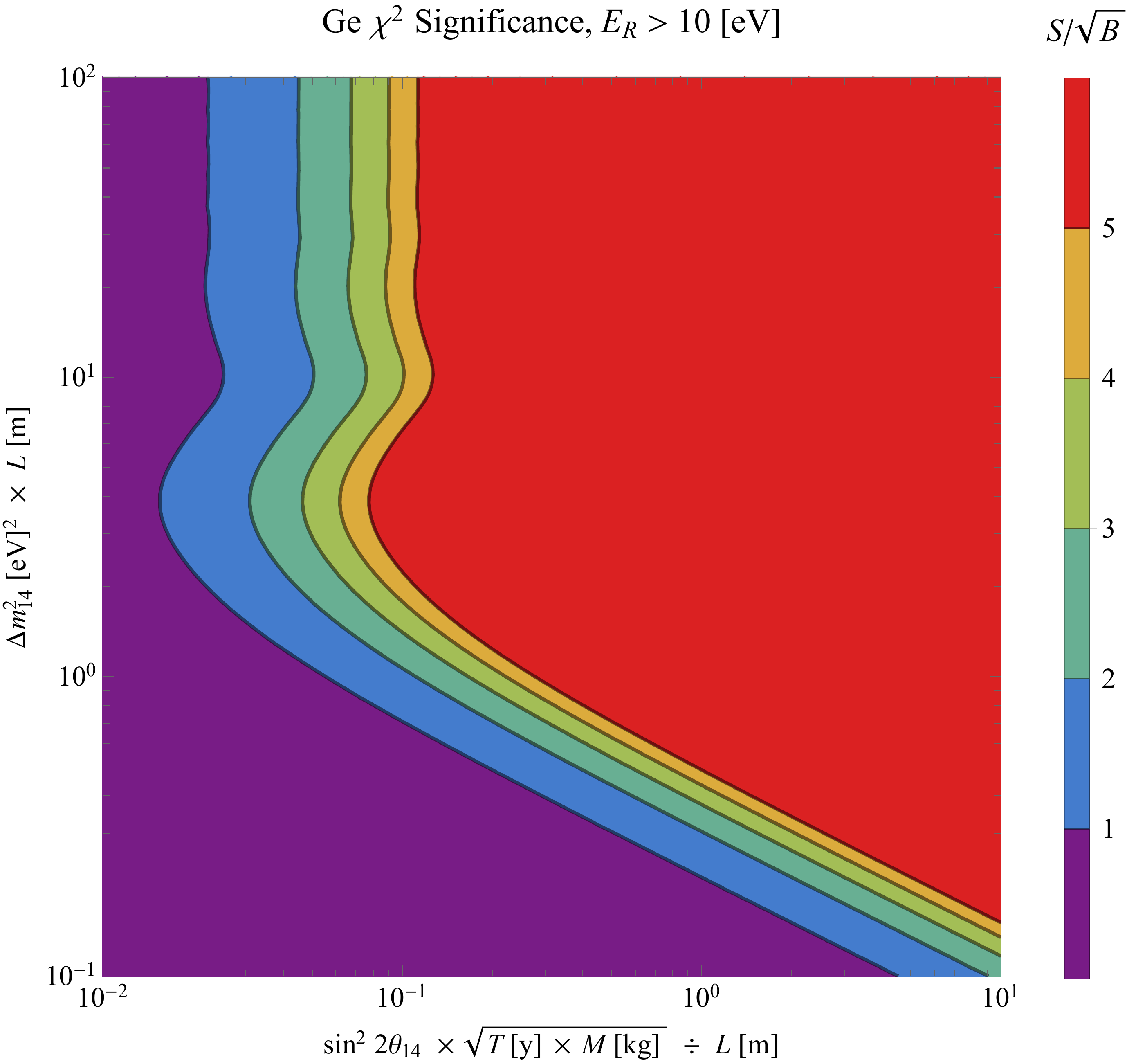}
\caption{
\footnotesize Figures depict contours in the $\chi^2$ statistical significance,
specifically $S/\sqrt{B}$, where the signal $S$ corresponds to an event deficit
attributable to oscillation with a sterile fourth generation neutrino
relative to the background expectation $B$ for
the SM electron anti-neutrino \cns{} event rate in \ger{}.
Scale invariant projections are made in terms of the mass gap $\Delta m_{14}^2~{\rm eV}^2$
and the amplitude $\sin^2 2 \theta_{14}$, as a product or quotient
with the experimental baseline $L$~m, respectively, and times the square root of exposure in the latter case.
At unit distance, exposure time and mass these factors drop out, and the axes may be read traditionally.
The detector recoil energy is unbinned in this analysis.
The left- and right-hand panel integrates all recoils above the more conservative and more
optimistic thresholds of 100 and 10~eV, respectively.
}
\label{fig:sig}
\end{figure}

The oscillated event expectation $N_{\rm Osc}^1$ in the $i^{\rm th}$ $E^i_{\rm R}$ recoil energy bin
is given in terms of the baseline SM expectation $N^i_{\rm Exp}$,
the oscillation amplitude $\sin^2 2\theta_{14}$, and the convolved deviation shape functional
$\gamma_i(\Delta m_{14}^2 L)$ of Eq.~(\ref{eq:gamma}).
\be
N^i_{\rm Osc} = N^i_{\rm Exp} \times \left\{ 1 \,-\, \sin^2 (2\theta_{14}) \, \gamma_i(\Delta m_{14}^2 L) \right\}
\label{eq:nosc}
\ee
In the absence of data, it is still quite possible to roughly estimate the sensitivity of a counting experiment
to deviations from the null result in a simplified manner. Referencing Eq.~(\ref{eq:nosc}), we construct a $\chi^2$
statistic comparing the deviation-squared of the oscillated signal $N_{\rm Osc}^i$ from the SM \cns{} expectation
(which is construed now as background)
to the associated statistical uncertainty $\sigma_i \sim \surd N_{\rm Exp}^i$ in this background rate.
We sum over $C$ observation channels, where the binning index $i$ momentarily performs double duty,
labeling both the targeted range of recoil energies and the detector location.
\begin{equation}
\chi^2 \,\equiv\, \sum_{i=1}^C \frac{(N^i_{\rm Osc}-N^i_{\rm Exp})^2}{N^i_{\rm Exp}}
\,=\,
\sin^4 2\theta_{14} \times \sum_{i=1}^C \gamma_i^2\, N^i_{\rm Exp}
\label{eq:chisq}
\end{equation}
We may (and will) further consider inclusion of projections for the unreduced non-\cns{} experimental backgrounds
(as described in Section~\ref{sec:errors}) into a unified event profile baseline, above which
the amplitude of sterile neutrino induced oscillation should be statistically visible.  We justify
neglect of systematic uncertainties on the shape and normalization of the unified background event
profile in two ways.  Firstly, as described in the prior section,
the amount of data collection required to suppress uncertainty in the
sideband calibration of non-\cns{} backgrounds has been demonstrated to be well within reach.
Secondly, as will be elaborated in the subsequent sections, a more complete analysis using
log-likelihood techniques with an oscillatory template hypothesis may be formulated to
induce cancellation of leading order uncertainties that do not present with an oscillating event profile.
Note further (for example) that a 1 year run will give on the order of $10^5$
recoils at one meter and the flux normalization would carry a 0.3\% (0.5\% for $E_R>100$ eV) uncertainty,
suggesting (in this case) that systematic flux uncertainties in reactor neutrino and neutron
backgrounds are subdominant for $\sin^22\theta \gtrsim 0.01$.

In the limit where many stochastically dispersed binning channels $C$ are sampled with an approximately uniform
distribution of expected counts $N_{\rm Exp}^i \simeq N_{\rm Tot}/C$, the value of Eq.~(\ref{eq:chisq})
will converge to $\chi^2 \to 3/8\, N_{\rm Tot}\, \sin^4 2\theta$, where the numerical coefficient represents a
fourth moment $\langle\,\sin^4\,\rangle = 3/8$ of the sinusoid embedded within $\gamma_i$.
The result is independent of the number of channels $C$, and is identical
to the scenario where samples are unbinned. This indicates that statistical significance of the deviation
declines in this scenario with the isolation of samples into multiple bins, because the fixed $\chi^2$ value is
then distributed over more degrees of freedom $C$. The result is readily understood, and is attributable to
the fact that the sign of $\gamma_i$ is always positive, i.e. the sterile neutrino always effects
a downward fluctuation in the event rate.

The $\chi^2$ significance of the oscillation-induced anti-neutrino deficit
relative to the statistical background at a single experimental baseline $L$,
and with no binning in the nuclear kinetic recoil,
is projected in FIGS.~(\ref{fig:sig}) as a function
of $\Delta m^2_{14}$ and $\sin^2 2\theta_{14}$.
Specifically, the exhibited contours correspond to integer values in the range
of 1 to 5 for the test statistic $S/\sqrt{B}$, interpreting $S$ as the event depletion
attributable to sterile neutrino oscillation effects, and $B$ as the unified SM \cns{} event expectation.
Results are presented for both conservative and optimistic nuclear recoil thresholds of 100 and 10~eV.
In order to manifest the generality and elegance of underlying scaling symmetries
with respect to distance $L$ and the exposure time $T$ and mass $M$, backgrounds
which do not respect the same scaling law are neglected in these figures.
As expected from Eq.~(\ref{eq:oscprob}) and FIG.~(\ref{fig:oscillation}),
observability is greatly diminished in the vertical axis whenever
$(\Delta m^2_{14}~{\rm eV}^2 \times L~[{\rm m}] \ll 1)$, as
there is insufficient phase evolution. Likewise, as suggested by Eq.~(\ref{eq:chisq}), observability
in the horizontal axis is hampered by reduction of the oscillation amplitude $\sin^2 2\theta_{14}$,
reduction of the exposure, or by elongation of the separation from core
(via geometric reduction in the neutrino flux as $N^i_{\rm Exp} \propto 1/L^2$).

\begin{figure}[b]
\includegraphics[width=3.2in]{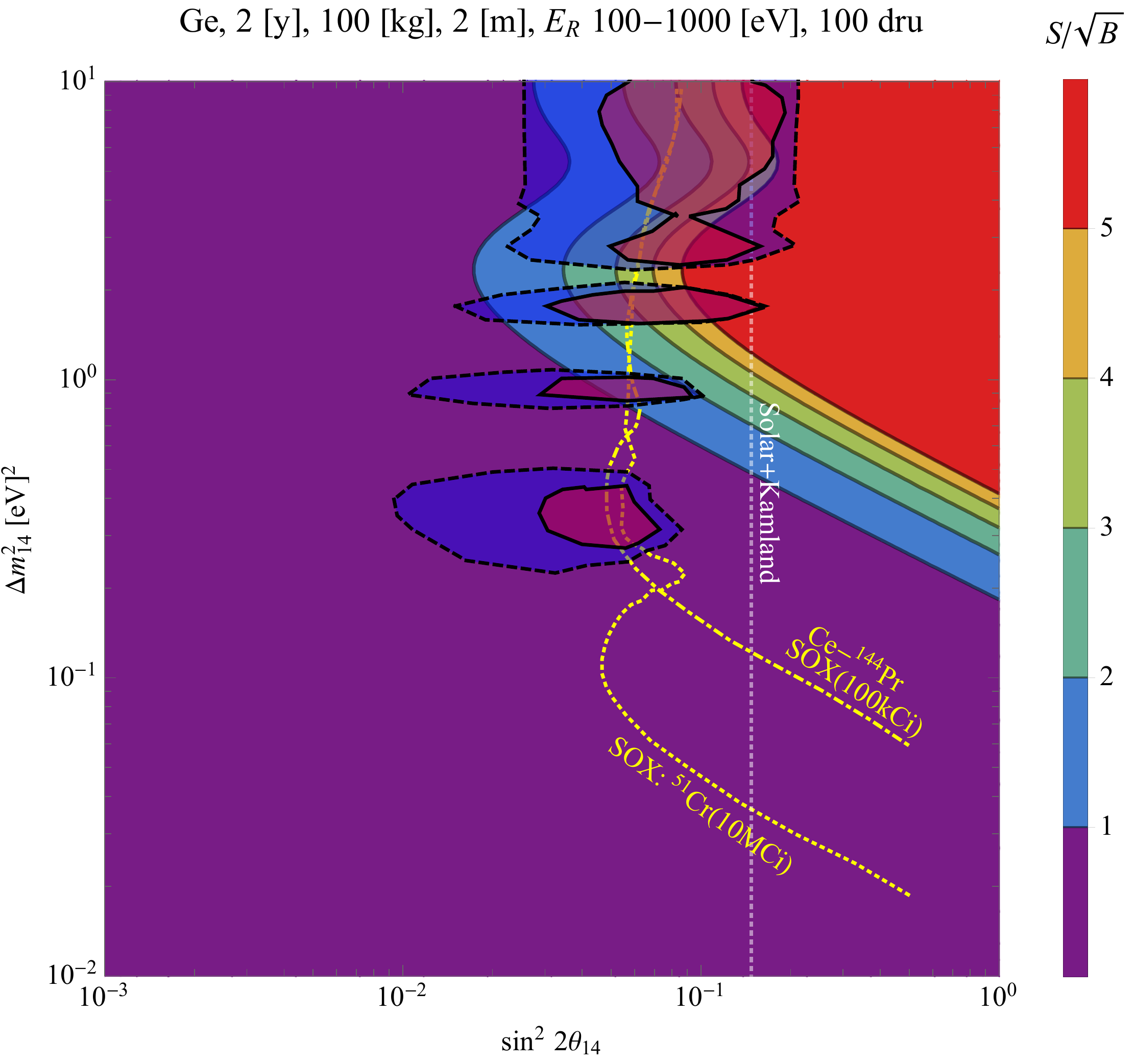}
\includegraphics[width=3.2in]{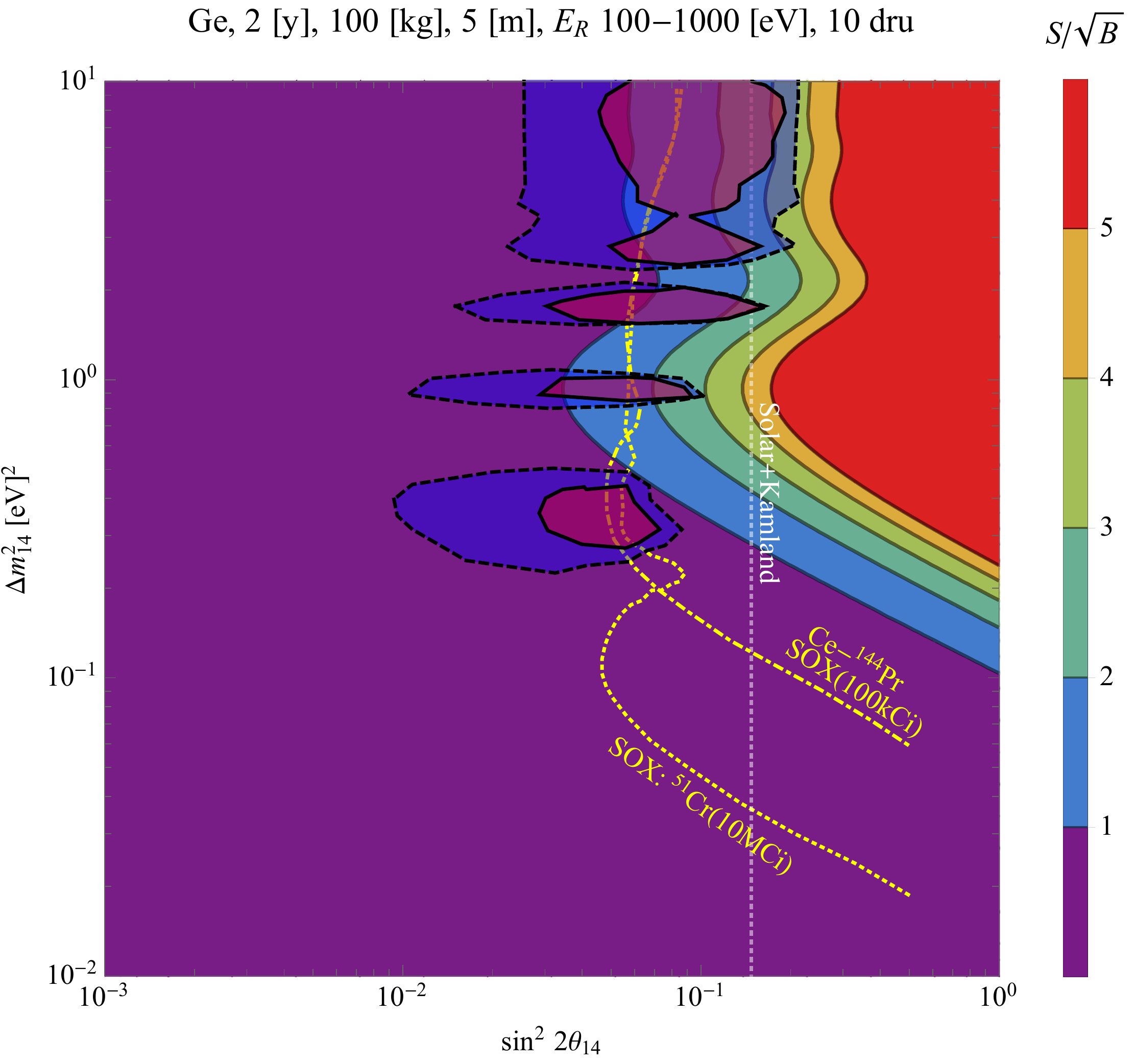}
\caption{\footnotesize Phase-1 prospective limit of $\nu_e-\nu_s$ mixing parameters with 100~kg Ge detector mass and 2
year effective exposure at a sample distance of 2~m (left) or 5~m (right) from the reactor.
A minimum recoil threshold of 100~eV is adopted.
Contours are shown in the $\chi^2$ statistical significance,
specifically $S/\sqrt{B}$, where the signal $S$ corresponds to the
oscillation event deficit and the background $B$ sums 
the SM electron anti-neutrino \cns{} event rate together with a postulate
for the unified reactor plus cosmic, etc. event rate.
This additional background component is modeled with
a strength of 100 or 10 dru at the near and far positions, respectively.
Global 95\% short-baseline (blue dashed) and $\nu_e$ disappearance (red solid)
fits~\cite{Kopp:2013vaa} and projected SOX~\cite{Borexino:2013xxa}
and Solar + Kamland~\cite{Billard:2014yka} limits are plotted for rough comparison.}
\label{fig:prospects-phase1}
\end{figure}

FIGS.~\ref{fig:prospects-phase1} show the specific sensitivity projected for a 100~kg payload with
a 2~yr exposure, using the phase-1 recoil sensitivity threshold of 10~eV.  Figures are presented
with both a 2 and 5~m separation between the detector and the reactor core.
Additionally, unified neutron, gamma, and cosmic (etc.) background event rates of 100 and
10~dru are folded into the analysis in the near and far positions, respectively.
For purposes of rough comparison (approaches to limit setting and handling of systematics are not identical),
global fit contours at 95\% confidence for short-baseline (blue dashed) and $\nu_e$ disappearance (red solid)
from Ref.~\cite{Kopp:2013vaa}, as well as projected limits from SOX~\cite{Borexino:2013xxa}
and Solar + Kamland~\cite{Billard:2014yka}, are overlaid.  As indicated, there is good sensitivity overlap
with the primary region of interest, although the described phase-1 scenario is limited with respect
to probing the low mass-gap parameter space. 

\begin{figure}[b]
\includegraphics[width=3.2in]{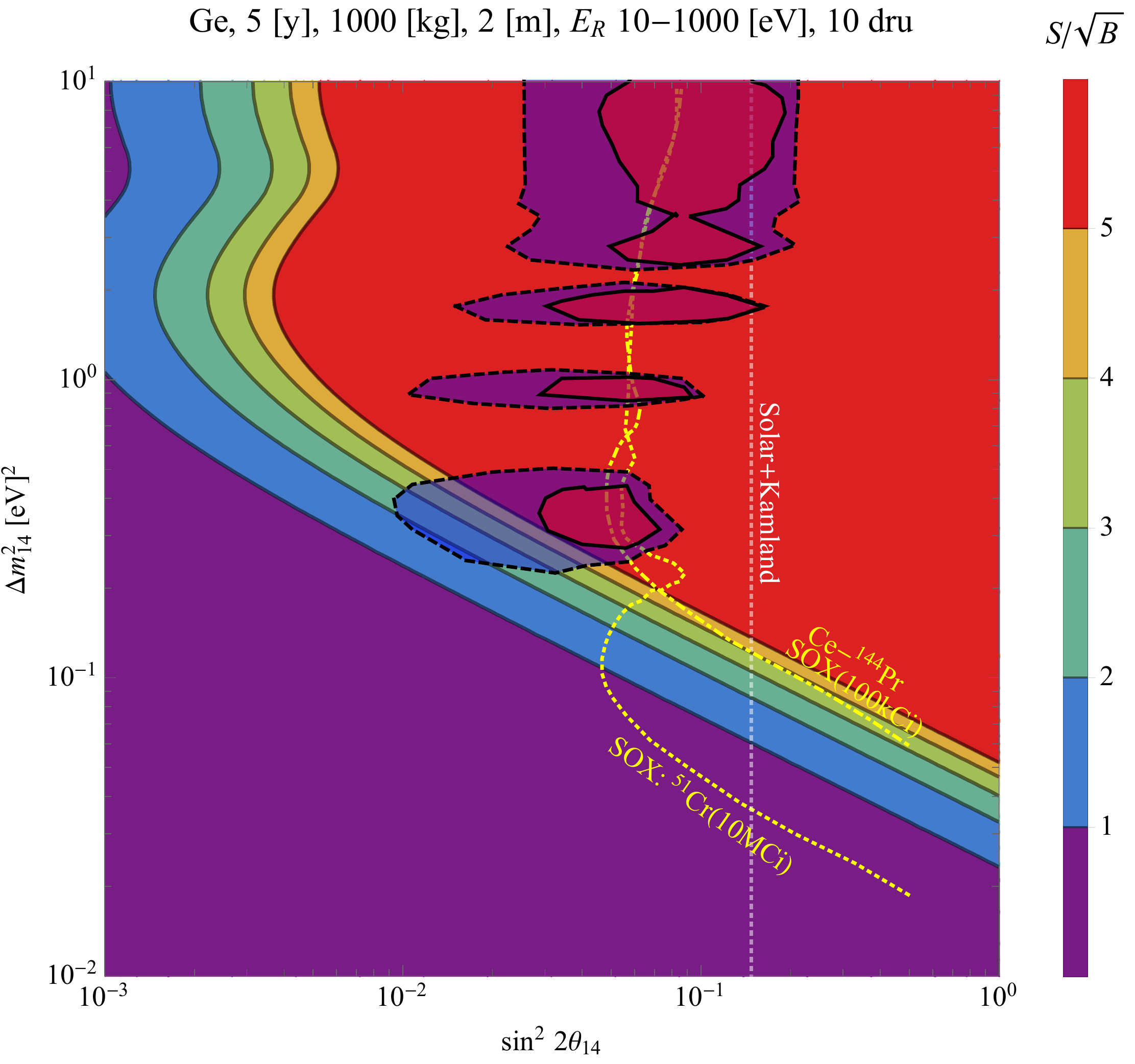}
\includegraphics[width=3.2in]{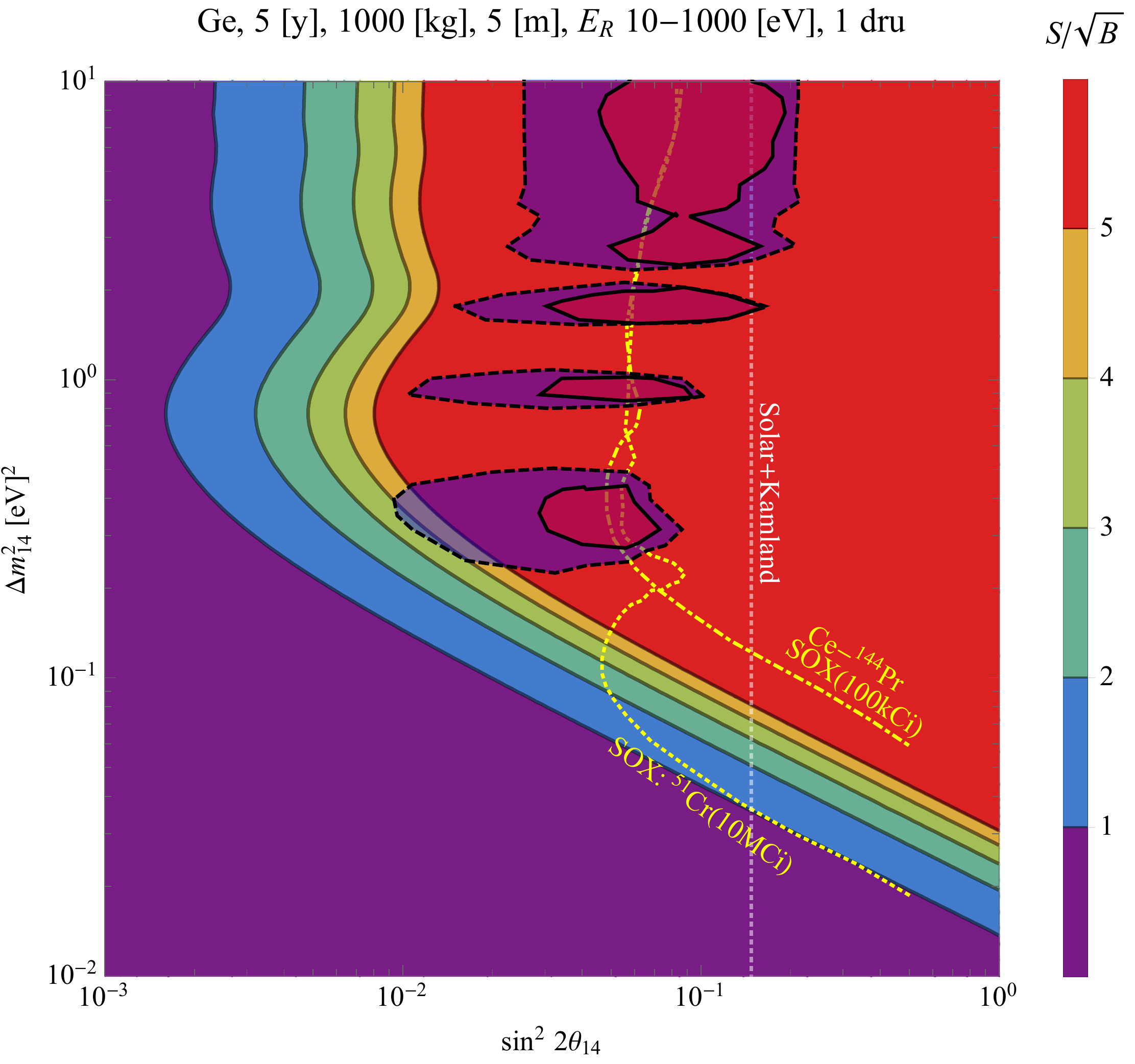}
\caption{\footnotesize Phase-2 prospective limit of $\nu_e-\nu_s$ mixing parameters with 1000~kg Ge detector mass
and 5 year effective exposure at a sample distance of 2~m (left) or 5~m (right) from the reactor.
A minimum recoil threshold of 10~eV is adopted.
The additional background component is modeled with
a strength of 10 or 1 dru at the near and far positions, respectively.
Contours and overlays are as described in FIGS.~\ref{fig:prospects-phase1}.}
\label{fig:prospects-phase2}
\end{figure}

FIGS.~\ref{fig:prospects-phase2} repeat the prior analysis with escalation to
a 1~ton payload and a 5~year exposure.  Simultaneously, we adopt for this phase-2 analysis,
an improved (speculative) recoil sensitivity threshold of 10~eV.  In keeping with this more
optimistic and forward-looking scenario, we also assume shielding improvements such that
background rate is limited to 10 and 1~dru at 2 and 5~meters, respectively.
The projected discovery contours for this second analysis
essentially covers the allowed space of $\Delta m^2_{14}$ and $\sin^2 2\theta_{14}$ values
associated with global fits to reactor and gallium experiments~\cite{Kopp:2013vaa},
and projected sensitivity to the $\bar{\nu}_e$-sterile mixing
is competitive with that of the SOX experiment~\cite{Borexino:2013xxa}.

If the reactor source and the expected SM background rate $N^i_{\rm Exp}$ are very well understood,
then the uniform depletion of events at long distance scales in the $(\gamma_i \Rightarrow 1/2)$ dispersion limit is
still quite visible. However, a constant deficit of this type is not intrinsically separable from systematic errors in the
event rate normalization. Again, the inherent advantage of the oscillation signal is its spatial structure, which
allows for the handy cancellation of systematics in the underlying event rate. On the other hand, this also implies that
the detector will then only be sensitive to oscillation wavelengths $\lambda$ on the order of the distance to core $L$.
The methodology appropriate to an analysis of this second type is developed subsequently. 

\section{Measurement of Parameters and Quantification of Statistical Bounds}
\label{sec:statistics}

When data is available, either of the experimental or Monte Carlo variety,
an analysis quite different than that described in Section~\ref{sec:sensitivity} may be conducted,
wherein the observed counts $N_{\rm Obs}^i$ in the $i^{\rm th}$ distance and/or energy
bin are compared to the expectation value $N^i_{\rm Osc}$, cf. Eq.~(\ref{eq:nosc}),
under an oscillation hypothesis template.
Allowing for an undetermined renormalization $\alpha \simeq 1$ of the overall event rate expectation,
the probability of observing a deviation of $N_{\rm Obs}^i$ events relative to the rate $\alpha N_{\rm
Osc}^i$
is normally distributed with respect to the bin uncertainty width $\sigma_i$.
\be
\mathcal{P}_i \equiv \mathcal{P}(N_{\rm Obs}^i) = \exp\left\{- \, \frac{{\left[ N_{\rm Obs}^i - \alpha N_{\rm Osc}^i \right]}^2}{\sigma_i^2}\right\}
\label{eq:pnorm}
\ee
Given definite values for the binned observation $N_{\rm Obs}^i$, expectation $N_{\rm Exp}^i$ and uncertainty $\sigma_i$,
we may invert the prior interpretation such that the likelihood
$\mathcal{L}_i \equiv \mathcal{L}(\alpha,\sin^2 2\theta,\Delta m^2)$ of the actual underlying global scale,
amplitude, and phase parameters aligning with some specified triplet of floating values is numerically proportional to
(or, conveniently, equivalent to) the Eq.~(\ref{eq:pnorm}) probability density $\mathcal{L}_i \equiv \mathcal{P}_i$.
The joint likelihood $\mathcal{L}_C = \prod_{i=1}^C \mathcal{L}_i$, which is a product of likelihoods for each of the $C$
binned observation channels, remains functionally dependent upon just three global degrees of freedom.
The negative-log $\widetilde{\mathcal{L}_C} \equiv - \ln \mathcal{L}_C$ of this quantity varies monotonically with
the original likelihood, and is minimized at the same point in parameter space at which the former is maximized; it
furthermore converts the product into a sum over $\chi^2$-type terms.
\be
\widetilde{\mathcal{L}_C} = \sum_{i=1}^C
\frac{{\left[ N_{\rm Obs}^i - \alpha N_{\rm Osc}^i \right]}^2}{\sigma_i^2}
\label{eq:like}
\ee
The joint log-likelihood $\mathcal{L}_C$ may be minimized numerically in order to fix each of
the parameters $\alpha$, $\sin^2 2\theta$ and $\Delta m^2$, while establishing error bounds
associated with deterioration of the fit as their values flow away from the optimization.
However, it is instructive and potentially useful to first semi-analytically reduce this expression
with regards to the scale renormalization $\alpha$.
Setting $\partial \widetilde{\mathcal{L}_C}/\partial \alpha = 0$ via Eq.~(\ref{eq:like}),
and referencing Eq.~(\ref{eq:nosc}) with the compact notation $\widehat{s} \equiv \sin^2 2\theta_{14}$, we get the following.
\be
\sum_{i=1}^C
\frac{{\left[ N_{\rm Obs}^i - \alpha \, N^i_{\rm Exp} \times \left\{ 1 \,-\, \widehat{s} \, \gamma_i \right\} \right]
\times N^i_{\rm Exp} \times \left\{ 1 \,-\, \widehat{s} \, \gamma_i \right\} }}{\sigma_i^2}
=0
\label{eq:optalpha}
\ee
This conditional does not, in general, lead to a compact closed-form expression for $\alpha$, and certainly not
one that is independent of the detailed structure of the shape functionals $\gamma_i$. However, certain well-motivated
approximations will induce a substantial simplification. Firstly, we will now restrict the interpretation of
Eqs.~(\ref{eq:like},\ref{eq:optalpha}) to a single binned energy range at a time, e.g. one of the TABLE~\ref{tab:expected}
aggregations, restricting the summation index $n$ to trace only over the range of distances $L_n$ from core.
Employing the Eq.~(\ref{eq:tsamples}) time allocation convention, the expected event count $N_{\rm Exp}^i$ in the
$i^{\rm th}$ energy bin is constant with respect to sampling done across positions;  this is an approximation,
with associated systematic error contributions, e.g. insomuch as it relies upon the uniform subtraction of
previously elaborated cosmogenic and related backgrounds that do not decline with distance-from-core.
Given this, it is reasonable to likewise stipulate an energy-binned error width $\sigma_i$ that
does not fluctuate with the position $n$. Finally, we will marginalize over the value of the
$\sin^2$ spatial oscillation shape $\gamma_i \to \langle\gamma_{i,n}\rangle = 1/2$.  This choice implies
that we must later be somewhat careful to elect a uniform distribution of spatial samples, with
respect the expected fluctuations of $\gamma_i(\Delta m^2 L)$ about the mean. However, that is
not difficult, as the distances scales $L_n$ most sensitive to the oscillatory signal component
at a given mass gap $\Delta m^2$ are precisely known.
With these modifications, Eq.~(\ref{eq:optalpha}) reduces to the following condition with respect to the
mean observed count per observation $\langle N_{\rm Obs}^i\rangle \equiv \sum_{n=0}^N N_{\rm Obs}^{i,n} \,/\, N$
(or the obvious modification thereof when $n$ is restricted to a subset
of $[\,0 , N\,]$ that ensures $\langle\gamma_{i,n}\rangle \simeq 1/2$~).
\be
\alpha = \frac{\langle N_{\rm Obs}^i \rangle}{ N_{\rm Exp}^i \times \left\{1 \,-\, \widehat{s}/2\right\}}
\label{eq:fixalpha}
\ee 
The interpretation of this result is straightforward; the midline of the expected oscillation depicted
in FIG.~(\ref{fig:oscillation}) has been renormalized to coincide with the mean of observations
over various experimental baselines $L_n$. This scaling induces the cancellation of leading systematic
errors associated, for example, with the reactor flux normalization.
Now, we will assign the uncertainty width $\sigma_i \to \surd\langle N_{\rm Osc}^i \rangle$
attending each observation in keeping with the dominant residual statistical fluctuation.
It is simultaneously convenient to define a pair of measures of the oscillatory deviation
from the mean, for the observed and expected signal, respectively.
\begin{align}
\Delta_{\rm Obs}^{i,n} &\,\equiv\, \frac{ N_{\rm Obs}^{i,n} }{\langle N_{\rm Obs}^i \rangle} \,-\, 1 \\
\Delta_{\rm Osc}^{i,n} &\,\equiv\, \widehat{s} \times \left\{ \frac{ 1 \,-\, 2\, \gamma_{i,n} }{ 2 \,- \,\widehat{s} } \right\}
\label{eq:delta}
\end{align} 
In terms of these factors, the joint log-likelihood referenced in Eq.~(\ref{eq:like})
for all observations in a given energy bin may be succinctly rephrased as follows,
where the subscript on $\widetilde{\mathcal{L}^i_2}$ references the pair of remaining
degrees of freedom in the template, namely $\sin^2 2\theta_{14}$ and $\Delta m^2_{14}$.
\be
\widetilde{\mathcal{L}^i_2} \,=\, \langle N_{\rm Obs}^i \rangle \times \sum_{n=0}^N
{\left\{ \Delta_{\rm Obs}^{i,n} - \Delta_{\rm Osc}^{i,n} \right\}}^2
\label{eq:like2}
\ee
The goodness of a fit to data may be quantified by the comparison of this factor to the
zeroth order fit $\widetilde{\mathcal{L}^i_0}$, where parameterization freedom has
been redundantly eliminated by simultaneous application of the SM
limits $\widehat{s} \to 0$ and $\gamma_i \to 1/2$, which imply $\Delta_{\rm Osc}^{i,n} \to 0$.
\be
\widetilde{\mathcal{L}^i_0} \,=\, \sum_{n=0}^N
\frac{{\left\{ N_{\rm Obs}^{i,n} - \langle N_{\rm Obs}^i \rangle  \right\}}^2}{\langle N_{\rm Obs}^i\rangle}
\,=\, (N+1)\times\frac{\sigma_i^2}{\langle N_{\rm Obs}^i\rangle}
\,\ge\, (N+1)
\label{eq:like0}
\ee
The expected variance $\sigma_i^2$ should be comparable to $\langle N_{\rm Obs}^i\rangle$ if it arises
from a purely statistical mechanism, but may be (substantially) larger if the signal has a substantial
oscillatory component that is not well modeled by the flat zeroth order template.
The result is expected to grow in proportion to the number of elected bins. There is
expected to be an advantage for $\widetilde{\mathcal{L}^i_2}$ over $\widetilde{\mathcal{L}^i_0}$
with respect to the inclusion of additional sample points if and only if the signal is
within the strongly oscillating central region of the $\Delta m^2 L$ space, cf.  FIG.~(\ref{fig:oscillation}).
Judicious preselection of the indices $n$ to be retained in each region of the likelihood
optimization should be performed in a manner consistent with
the $\langle\gamma_{i,n}\rangle \simeq 1/2$ condition.

Wilk's theorem states that twice the difference of negative-log-likelihoods
for nestable model templates is approximately $\chi^2_D$ distributed,
with degrees of freedom $D$ equal to the difference in number of optimized parameters.
Specifically, the criterion for a significant improvement, at a type-I error level $p$,
for the 2-parameter template fit over the constrained fit is as follows, where
the cumulative distribution function (CDF) represents the fraction of parameter space bounded
within a multi-dimensional (typically Gaussian) integration out to some ``radius'' $\chi$.
\be
\widetilde{\mathcal{L}_2^i}
\,\le\,
\widetilde{\mathcal{L}_0^i} -
{\rm CDF}^{-1}\big(\chi^2_2,1-p\big)/2
\label{eq:wilks}
\ee
The inverse CDF is simply the $\chi^2_D$ boundary value in $D$ dimensions
for which a fraction $p$ of possible outcomes would be considered more extreme.
For example, with $D=2$, the inverse CDF $\chi^2$-values for confidences levels $(1-p)$
corresponding to $\{68,95,99.7\}\%$, i.e. $\{1,2,3\}\times\sigma_i$, are $\{2.3,6.2,11.8\}$,
respectively.
This is a one-tailed test, insomuch as the negative-log-likelihood of the
parameterized fit must necessarily meet or exceed that of the constrained fit.
Eq.~(\ref{eq:wilks}) may alternatively be inverted to solve for the $p$-value,
which may be converted into an equivalent significance multiple $\mathfrak{N}\times\sigma_i$
of the one-dimensional Gaussian standard deviation.
\begin{align}
p \,&=\, 1 - {\rm CDF}\big(\,\chi_2^2,\,
2\times\big[\widetilde{\mathcal{L}_0^i} - \widetilde{\mathcal{L}_2^i}\big]\,\big) \\
\mathfrak{N} \,&=\, \sqrt{\,{\rm CDF}^{-1}\big(\chi^2_1,1-p\big)\,}
\label{eq:wilksinv}
\end{align}
If the oscillation template is demonstrated to be significantly superior to the null template
in a given energy bin $i$, then uncertainty bounds on the optimized parameters
$\sin^2 2\theta_i$ and $\Delta m^2_{i}$ may be established via the computation of likelihood profiles.
In turn, the one-dimensional likelihood $\widetilde{\mathcal{L}^{i,j}_1}(q)$ of the
$j^{th}$ parameter fit is sampled for values $q^j$ adjacent to the optimization $q_0^j$,
while all conjugate parameters are re-optimized. Then, the cutoff at a level $p$ corresponds to
the value of $q^i$ for which $\widetilde{\mathcal{L}^i_1}(q^j) -
\widetilde{\mathcal{L}^i_1}(q_0^j)$ is
equal to ${\rm CDF}^{-1}\big(\chi^2_1,1-p\big)/2$. Results for each parameter may then be
consolidated across independent energy bins in order to establish combined limits.

It is expected that the sensitivity delivered by this type of analysis (or similarly
for the generalized approach sketched in the final section), where one takes full
advantage of the additional handles provided by the oscillatory event profile (including 
embedded correlations across distance and recoil energy bins), could substantially exceed the
somewhat na\"{\i}ve sensitivity estimates computed in Section~\ref{sec:sensitivity}
by comparing the unbinned signal amplitude to the scale of fluctuations in the
unified \cns{} observation rate (including non-\cns{} backgrounds) at a single experimental baseline. 
It is of great interest to the authors to undertake a full trial analysis
of simulated data with Monte Carlo event population, folding in realistic backgrounds and experimental
systematics, applying the statistical techniques described in the present work to the extraction
of signal and the setting of limits.  Only by such means is it possible to
extract information encoded in the variable experimental baseline and spectral decomposition
of the recoil energy spectrum while deeply suppressing systematic uncertainties in the
underlying non-oscillatory event profile. We briefly lay out the formalism in Section~\ref{sec:statistics2}.
This exercise, which would require optimization of the realistic measurement plan and exceeds the 
scope of the current document, is reserved for future work.

\section{Backgrounds and a generalized fit}
\label{sec:statistics2}

The recoil measurement will face several backgrounds that need to be taken into account.
In this section we discuss a general formalism of the likelihood fit, and a proof-of-principle
treatment of the major background categories listed in Section~\ref{sec:errors}.
To build a general template for the recoil event rate, let us formulate the total
recoil event number at each detector location as
\be
N_i=T_i \int_{E_{R,i}^{min}}^{E_{R,i}^{max}} dE_{\rm R} \left(\frac{L_0}{L_{i}}\right)^2
\left(\alpha f_{\nu}(E_{\rm R})
\frac{dN_\nu}{dE_{\rm R}} + \beta f_{n}(E_{\rm R}) \frac{dN_n}{dE_{\rm R}}\right) + \gamma f_{\mu n}(E_{\rm R}) \frac{dN_{\mu n}}{dE_{\rm R}},
\label{eq:Ni}
\ee
where $i$ is the index for different detector locations at a distance $L_i$ from the reactor core, $T_i$
is the time exposure, $L_0$ is a reference distance. $E_{\rm R}$ is the recoil energy. $dN/dE_{r}$
represents the flux from reactor neutrino recoils ($\frac{dN_\nu}{dE_{\rm R}}$), the reactor
gamma-ray induced (secondary neutron) events ($\frac{dN_\gamma}{dE_{\rm R}}$), 
and the cosmic muon induced neutrons ($\frac{dN_{\mu n}}{dE_{\rm R}}$)
which have penetrated through the cement overburden. The $f$ function denotes the energy-dependent detection
efficiencies. We ignore the `direct' reactor gamma-ray recoils,
which can be efficiently eliminated by shielding.

$\{\alpha,\beta,\gamma\}$ are a set of nuisance parameters that allow each component to float within their
normalization uncertainties $\{\Delta\alpha,\Delta\beta,\Delta\gamma\}$. By including active-sterile oscillations,
the reactor neutrino events will depend on sterile neutrino mixing parameters $\{\Delta m^2, \sin^22\theta\}$,
\be 
N_i =N_i(\{\Delta m^2, \sin^22\theta\};\{\alpha,\beta,\gamma\}).
\ee
The relative sizes between the oscillation signal and the normalization uncertainties play an important
role in the fitting strategy.
For a large sterile mixing angle, e.g. greater than the uncertainty on the reactor flux 
$\sin^22\theta > \Delta \alpha$, $\nu$ recoils yield a 
deviation (compared to the non-oscillation flux) no less than half of
the $\sin^22\theta$ amplitude over the entire energy range, and it is desirable to 
combine the data at all recoil energy to enhance likelihood significance.

However, when $\sin^22\theta$ is comparable or less than the reactor flux uncertainty, the flattened
half-amplitude-decrease from sterile neutrino oscillation becomes indistinguishable from a
overall fluctuation in the normalization of the reactor neutrino flux. In this case it can be more 
desirable to focus on the range of $E_{\rm R}$ where the oscillation has not flattened 
from dispersion (see the central range of FIG~\ref{fig:oscillation}).


With the estimated 100 dru (10 dru) background rates, a high-statistics calibration in the side-band can determine
the leading background normalizations to be less than 1 percent accuracy in days (weeks), by running a 
calibration fit for the nuisance parameters,
\be 
 \{\alpha_0,\beta_0,\gamma_0\} \text{  minimizes  } \sum_i \frac{[N_i(\{0,0\};\{\alpha,\beta,\gamma\})-N_i^{obs}]^2}{N_i^{obs}+sys^2},
\ee
where null sterile mixing parameters are assumed for calculating exclusion limits. Here the systematics
will include any uncertainty other than the fluctuations represented by
$\{\alpha,\beta,\gamma\}$. Next we can perform a fit to the sterile mixing parameters,
\be 
\chi^2 =\sum_i \frac{[N_i(\{\Delta m^2, \sin^22\theta\};\{\alpha,\beta,\gamma\})-N_i^{obs}]^2}{N_i^{obs}+sys^2} +
\frac{(\alpha-\alpha_0)^2}{(\Delta\alpha)^2} + \frac{(\beta-\beta_0)^2}{(\Delta\beta)^2} + \frac{(\gamma-\gamma_0)^2}{(\Delta\gamma)^2}
\ee

At each $\{\Delta m^2, \sin^22\theta\}$ point, $\{\alpha,\beta,\gamma\}$ are marginalized over to
obtain the best fit $\chi^2$ value. The additional nuisance parameter terms take care of the effect of
globally shifting the flux normalizations, and leave the number of degrees of freedom unchanged.
In the statistics-dominated limit, the nuisance parameters will be tied closely to their central values,
and the above formula reduces to Eq.~\ref{eq:chisq}.



The variation in energy and spatial distribution of backgrounds require the calibration process to be carried out
at each reactor-detector distance of measurement. Optimization between observation time and detector location
needs to be done in accordance with the total experimental time span. Well-placed cuts can also improve the signal 
to background ratio, for instance, requiring $E_R <200 $eV can efficiently control cosmic-muon induced recoils, 
whose flat energy distribution weighs more on higher $E_R$ in comparison to that of signal recoil events.

\section{Conclusions and Summary}
\label{sec:conclusions}

In this paper we have discussed a variable multi-meter-scale baseline measurement of active-sterile
neutrino oscillations in collaboration with the Nuclear Science Center at Texas A{\&}M University. Our
motivation is to test for the presence of a 4th generation sterile neutrino, which has been hinted at by several
experiments~\cite{Kaether:2010ag,Abdurashitov:2009tn,Giunti:2006bj,Giunti:2010zu,Mention:2011rk}.
A 4th generation sterile neutrino at the mass scale $\sim$ eV can be accommodated cosmologically, and is
consistent with Solar neutrino data.

We have shown that the adjustable reactor-detector distance gives our measurement scheme a great
advantage to sample different areas of the mixing parameter space, and that limits are competitive with projected
sensitivities from SOX~\cite{Borexino:2013xxa} and from detection of Solar
neutrinos~\cite{Billard:2014yka} with sufficient exposure.  A distance as near as 2-3~m emphasizes sensitivity to the
mixing angle, and farther positions provide improved reach for a small 4th generation neutrino
mass-square difference. Measurements at 2 and 5 meters, with a total 100 Kg Ge detector mass over an
effective exposure of 2 years probes a large fraction of the $\Delta m^2$ parameter space that is allowed by
short-baseline experiments.
With a conservative recoil energy threshold of 100~eV
we project an estimated sensitivity to first/fourth neutrino
oscillation with a mass gap $\Delta m^2 \sim 1 \, {\rm eV}^2$ at an
amplitude $\sin^2 2\theta \sim 10^{-1}$, or $\Delta m^2 \sim 0.2 \, {\rm eV}^2$ at unit amplitude.

Depending on the performance of the phase-1 experiment, a phase-2 experiment with a 1000 kg payload may be
proposed, to operate over a 5 year period to provide the maximum sensitivity achievable
from the reactor experiment at TAMU.
Such a next-generation experiment would be able to fully utilize and benefit from rapid technological
progress in the reduction of minimal recoil sensitivities (proceeding in parallel for applications
to low mass dark matter search experiments), which may conceivably be extended down to the 10~eV order.
In this case, it is possible to probe more than an additional order of magnitude in amplitude.
Such a second phase experiment might additionally be situated at an alternate site with
higher neutrino flux for the ultimate sensitivity to sterile neutrinos.

We have presented results accounting for the inclusion of reactor and cosmogenic backgrounds and
statistical uncertainties on the total experimental event rate.
We have provided an estimate for systematics, which include the anti-neutrino flux normalization,
and uncertainties in the other reactor and non-reactor backgrounds.
We have shown that spatial variation of the signal response provides an intrinsic mechanism for the reduction of systematics,
and have discussed methods for parameter estimation in the presence of both statistical and systematic uncertainties.

Neutrino mass provides the best evidence for physics beyond the Standard Model. Independent of the
aforementioned experiments that may hint at the presence of an additional sterile neutrino, this
provides clear motivation for us to understand the behavior of neutrinos over all accessible interaction
channels and energy scales. The experiment that we describe provides a unique channel to probe neutrino
interactions, and a promising method to search for physics beyond the Standard Model.

\section*{Acknowledgements} 
The authors gratefully acknowledge
contributions to the understanding of reactor normalization by Jan Vermaak,
additional contributions to the characterization of event backgrounds by Andrew Jastram and Will Flanagan,
and very helpful discussions with Enectal\'{\i} Figueroa-Feliciano.
BD acknowledges support from DOE Grant DE-FG02-13ER42020.
YG and AK acknowledge support from the Mitchell Institute for Fundamental Physics and Astronomy.
RM acknowledges very useful discussions with Rusty Harris.
LES acknowledges support from NSF grant PHY-1522717.
JWW acknowledges support from NSF grant PHY-1521105 and the Mitchell Institute for Fundamental Physics and Astronomy.

\bibliography{bibliography}

\end{document}